\documentclass[iop]{emulateapj}
\usepackage{threeparttable}
\usepackage{tabularx}
\usepackage{graphicx}
\usepackage{epstopdf}
\usepackage{graphics}
\usepackage{color}           
\usepackage{url}             

\shorttitle{Magnetic Energy Coupling}
\shortauthors{Orange et al.}

\begin{document}

\title{Solar Atmospheric Magnetic Energy Coupling: Broad Plasma Conditions and Spectrum Regimes }

\author{N. Brice Orange$^{1,2}$, David L. Chesny$^{1,3}$, Bruce Gendre$^{2,4,5}$, David C. Morris$^{2,4}$, and Hakeem M. Oluseyi$^{3}$}

\affil{$^1$OrangeWave Innovative Science, LLC, Moncks Corner, SC 29461}
\affil{$^2$Etelman Observatory, St. Thomas, United States Virgin Islands 00802}
\affil{$^3$Department of Physics \& Space Sciences, Florida Institute of Technology, Melbourne, FL  32901}
\affil{$^4$College of Science and Math, University of Virgin Islands, St. Thomas, United States Virgin Islands 00802}
\affil{$^5$Laboratoire ARTEMIS, Universit\'{e} C\^{o}te d'Azur, Observatoire\'{e} C\^{o}te d'Azur, CNRS, \\
\, boulevard de l'Observatoire, BP 34229, F-06304 Nice Cedex 04, France}

\begin{abstract}
Solar variability investigations that include magnetic energy coupling are paramount to solving many key solar/stellar physics problems, particularly for understanding the temporal variability of magnetic energy redistribution and heating processes. Using three years of observations from the {\it Solar Dynamics Observatory's} Atmospheric Imaging Assembly and Heliosemic Magnetic Imager; radiative and magnetic fluxes were measured from gross features and at full-disk scales, respectively. Magnetic energy coupling analyses support radiative flux descriptions via a plasma heating connectivity of dominant (magnetic) and diffuse
components, specifically of the predominantly closed field corona. Our work shows that this relationship favors an energetic redistribution efficiency across large temperature gradients, and potentially sheds light on the long withstanding issue of diffuse unresolved low corona emission. The intimacy of magnetic energy redistribution and plasma conditions revealed by this work holds significant insight for the field of stellar physics, as we have provided possible means for probing distant sources in currently limited and/or undetectable radiation distributions.
\end{abstract}

\section{Introduction}\label{sec:MCEIntro}

The Sun's atmosphere exists in two phases; one that is magnetically confined near the solar surface and one that consists of the extended atmosphere that interfaces with and comprises the solar wind. The solar atmosphere, observed on the disk and above the limb, can be divided into three distinct regions: active regions (ARs); regions of ``quiet" Sun (QS); and coronal holes (CH), i.e., gross feature classes. It has been established that the redistribution of magnetic energy appears to dominate the heating of the corona (e.g., \citealt{Klimchuk2015RSPTA}), but the mechanisms responsible for and heights at which plasma heating occurs remain outstanding puzzles.

Solar atmospheric heating of plasmas to coronal temperatures ($\log T$\,$\ge$\,6.0) is believed to result from the dissipation of magnetic free (i.e., via reconnection events) or wave energy; i.e., such energy conversion events lead to bundles of nanoflare heated loop strands \citep{Parker1963ApJS}. However, emerging evidence is challenging the standard coronal heating model, e.g., fast transition region (TR; 4.9\,$\le$\,$\log T$\,$\le$\,6.0) upflows \citep{Tripathietal2012ApJ,Orangeetal2013ApJ}, and strongly peaked active region core emission measure distributions \citep{Warrenetal2012ApJ}.

Throughout the last few decades, extensive work has been carried out on magnetically confined structures (e.g., \citealt{AschwandenSchrijver2002ApJS}; \citealt{Spadaroetal2006ApJ}; \citealt{Mackayetal2010SSRv} \citealt{Orangeetal2013ApJ}; \citealt{Chesnyetal2013ApJ}). These works, mainly in relation to the corona, have greatly influenced and enhanced our understanding of solar atmospheric heating (e.g., \citealt{AschwandenNightingale2005ApJ}), and revealed that both steady-state (e.g., \citealt{Winebargeretal2011ApJ}) and impulsive heating contribute to their generation (e.g., \citealt{ViallKlimchuk2012ApJ}). Basal heating of cooler atmospheric layers has been implicated as the source, and origin of the solar wind, respectively (e.g., \citealt{Cranmer2012SSRv}; \citealt{McIntoshetal2013arXiv}), which emanates from open field magnetic structures (e.g., \citealt{Lietal2012RAA}). Though investigations have sought the existence of a self-similar magnetically open and closed field heating mechanism (e.g., \citealt{LeeMagara2014PASJ}; \citealt{CheGoldstein2014ApJ}), little support exists for such \citep{Klimchuk2014arXiv}.

Key in pinning down a single dominant solar/stellar atmospheric heating mechanism of closed magnetic field structures, is the linear relationship of coronal X-ray luminosity to unsigned magnetic flux \citep{Pevtsovetal2003ApJ}. These results are supported by evidence of self-organized criticality (SOC; \citealt{Baketal1987PhRvL}), where heating events result from non-linear processes over broad spatial scales (e.g., \citealt{LuHamilton1991ApJ}; \citealt{Oluseyietal1999ApJ527-992O}). Moreover,   \citet{Alvarado-Gomez2016AA} have recently presented evidence from numerical simulations of the Sun and other cool main sequence stars of an extension of the  X-ray luminosity to unsigned magnetic flux relationship to the extreme ultra-violet (EUV). Therefore, it is of distinct interest to investigate if observational evidence supports an extension of linear radiative to magnetic coupling descriptions across previously unexplored electromagnetic spectrum regimes (i.e., visible, ultra-violet (UV), far UV (FUV), EUV, etc.), as well as temperature regimes (i.e., photospheric through coronal), multiple epochs of solar activity, and comparisons between large scale open and closed magnetic field structures (i.e., CH versus QS, etc.) remains unexplored.

Constraints on plausible heating mechanism(s) (e.g., \citealt{Mandrinietal2000ApJ}) can be ascertained from energetic coupling investigations of radiative and magnetic flux (e.g., \citealt{FludraIreland2003AA}). That is, observed intensities are dependent on thermodynamic distributions, subsequently governed by heating rates (e.g., \citealt{FludraIreland2003AA}; \citealt{WarrenWinebarger2006ApJ}). Importantly, the established magnetic field strength's role in heating models indicates that much stands to be learned of heating processes, and possibly variations thereof, via gross feature class comparison studies, considering that large thermodynamic gradients (e.g., \citealt{ODwyeretal2010AA}) and starkly differing magnetic field geometries (e.g., \citealt{Orangeetal2015ApJ}) should prevail between these features.

ARs are composed of the hottest and densest plasmas (e.g., \citealt{DelZannaetal2015AA}) across large temperature gradients, and hence, are the most luminous in the FUV, EUV, and soft X-ray. Of interest to ARs is that the most highly energetic transient phenomena in the solar atmosphere, e.g., flares (FL), predominantly occur in their cores, i.e., ``inter moss" regions, where plasma of $\log T$\,$>$\,6.3 resides \citep{Warrenetal2010ApJ,DelZannaetal2015AA} and densities exceed $\approx$\,10$^{10}$ cm$^{-3}$ \citep{ODwyeretal2010AA,ODwyeretal2011AA}.  The heating of ARs and their core (ARC) loops is a matter of much debate. Observations of the loops favor both low-frequency, i.e., stable high temperature emission from plasma heating rates much larger than cooling time scales, as well as high-frequency, i.e., shallow temperature gradients from heating rates less than cooling time scales (e.g., {\bf \citealt{Tripathietal2011ApJ}}; \citealt{Winebargetal2013ApJ}; see references therein). As other gross features and cooler atmospheric layers commonly indicate magnetically confined structures far from equilibrium, i.e., characterized by narrow temperature distributions (e.g., \citealt{AschwandenNightingale2005ApJ,Warrenetal2008ApJ685,Hansteenetal2014Sci}), it is apparent that their comparison to ARCs, across large electromagnetic spectrum regimes, are useful for deciphering the nature of plasma heating and its rates.

\begin{figure*}[!t]
\begin{center}
 \includegraphics[scale=0.23]{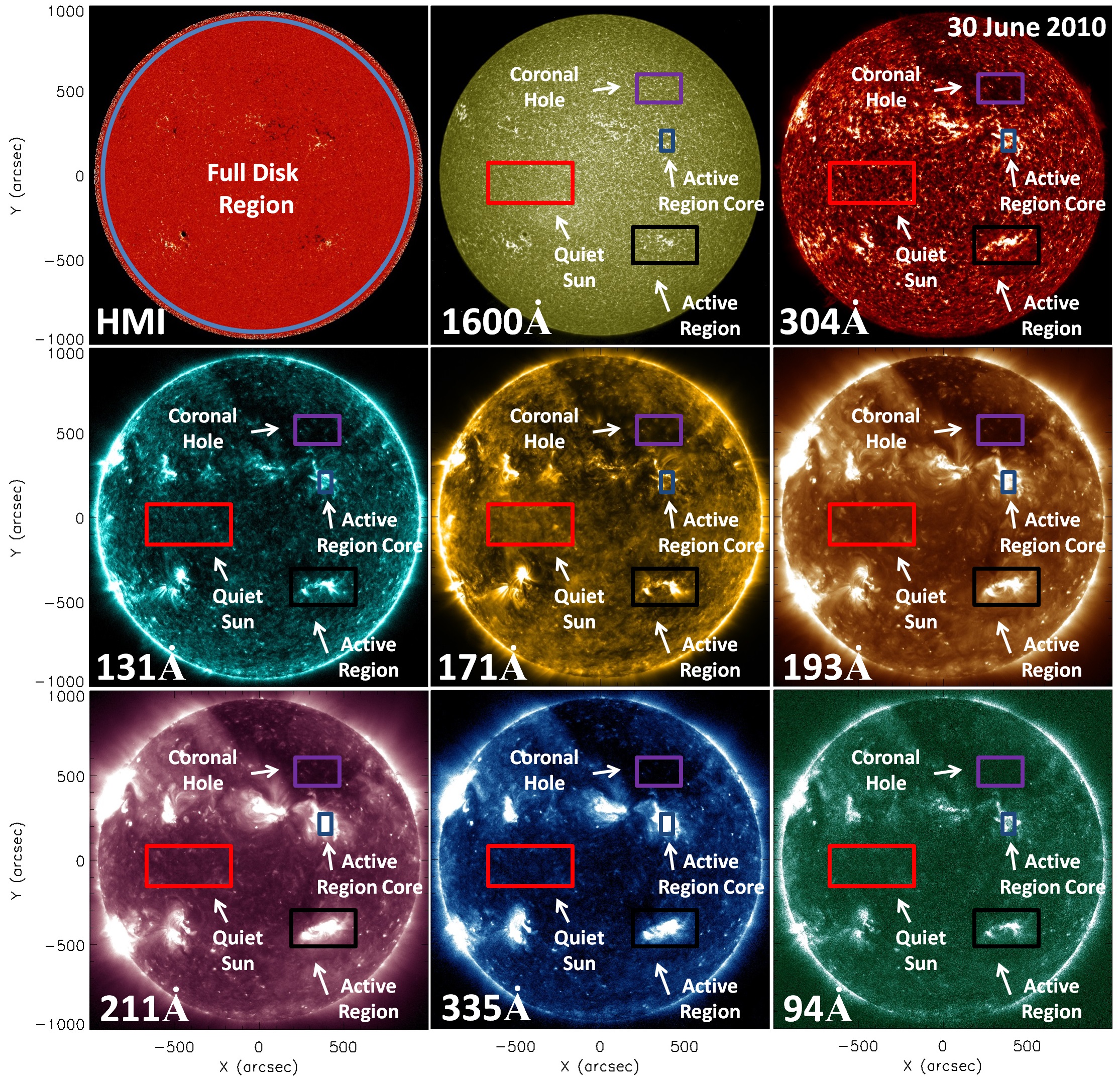}
 \caption{From left to right and top to bottom, respectively, HMI LOS magnetogram, and AIA 1600\,{\AA}, 304\,{\AA}, 131\,{\AA}, 171\,{\AA}, 193\,{\AA}, 211\,{\AA}, 335\,{\AA}, and 94\,{\AA} radiative images, respectively, observed 30 June 2010. Note, on HMI the circle (blue) indicates the region representing 95\% of the solar disk utilized to study the full disk feature reported on herein, while on each AIA radiative image examples of each of the other gross feature classes analyzed herein have been identified.}
\label{fig:SampleFeature}
\end{center}
\end{figure*}

In relation to the above presentation, and specifically to our goal of seeking a possible extension of the X-ray radiative to magnetic coupling description of \cite{Pevtsovetal2003ApJ} across broad electromagnetic spectrum regimes in the presence of large open and closed magnetic field structures, the remainder of this paper is organized as follows. Observational data processing and analysis of gross solar atmospheric feature classes (i.e., CHs, QS, ARs, and ARCs), as well as at full-disk (FD) scales are presented in Section~\ref{sec:MEC_ObsAnalysis}. Within Section~\ref{sec:MEC_results} we present radiative versus magnetic energy measurements (Section~\ref{sec:RadVsMag_TypWay}) and their linear energetic coupling descriptions, with and without feature dependence (Section~\ref{sec:RadVsMag_LinearPLaw} and ~\ref{sec:RadVsMag_BrokePLaw}, respectively). Section~\ref{sec:CoronalHeatingApp} presents a general coronal heating theory based on the compilation of our magnetic energy coupling analyses, and our conclusions are provided in Section~\ref{sec:MEC_SummaryConclusion}, respectively.

\section{Observations}\label{sec:MEC_ObsAnalysis}

Observational data was obtained from SDO's Atmospheric Imaging Assembly (AIA; \citeauthor{Lemenetal2012SolPh} \citeyear{Lemenetal2012SolPh}) and  Heliosemic Magnetic Imager (HMI; \citeauthor{Schouetal2012SoPh} \citeyear{Schouetal2012SoPh}) at approximately 3 -- 5 day intervals from May 2010 through July 2013. AIA data consisted of the following ten passbands: 94\,{\AA}, 131\,{\AA}, 171\,{\AA}, 193\,{\AA}, 211\,{\AA}, 304\,{\AA}, 335\,{\AA}, 1600\,{\AA}, 1700\,{\AA}, and 4500\,{\AA}, which image the Sun's full disk approximately every 12 s, with the exception of 4500\,{\AA} which observes at a typical cadence of $\approx$\,30 min. These bands observe solar plasma from photospheric to coronal temperatures at a pixel size of $\approx$\,0.6 arcsec pixel$^{-1}$. The HMI data are images of the full disk line-of-sight (LOS) magnetic field with a cadence of 45 s, and spatial resolution of $\approx$\,0.5 arcsec pixel$^{-1}$. AIA and HMI passband images were pre-processed using standard {\sf Solar SoftWare} (SSW), corrected for solar rotation effects, and co-aligned using the techniques of \citet{Orangeetal2014SoPh1901O}. {\bf Note that rotation effects between passbands were negligible through using observational time differences below AIA's thermal jitter motion ($\approx$\,0.$\arcsec$3; \citealt{Aschwandenetal2011SolPh}; \citealt{Lemenetal2012SolPh}; \citealt{Orangeetal2014SoPh1901O}), and that the applied alignment technique centered on utilizing the 1700\,{\AA} observations as the fiducial passband to which all others were co-registered.}

AIA 193\,{\AA} images, per observational date, were used to select two CH, QS, ARs and ARCs (e.g., Figure~\ref{fig:SampleFeature}). For each selected feature all AIA passband and HMI LOS magnetogram data were aggregated, and the typical radiative and unsigned magnetic fluxes measured, respectively. Errors were propagated using a summation of photon counting statistics, and the standard error on the mean. We note here; all investigations of solar atmospheric thermal to magnetic energy coupling in this work are carried out via the common approximation that energy flux is proportional to ``data numbers" (DNs; e.g., \citealt{Wolfsonetal2000ApJ}; \citealt{Benevolenskayaetal2002ApJ}), i.e., AIA data are not calibrated to physical units. We also recognize that no objective method exists in relation to identifying gross features such as CHs, QS, ARs, and ARCs. As such, spurious feature signal overlap should be expected within analyzed radiative distributions. However, as observed in Figure~\ref{fig:FvsPhi}, particularly in the 193\,{\AA} panel (i.e., passband used to identify features), we find confidence in the implemented selection methodology. Specifically and importantly, in being one which provided radiatively differing gross feature samples whose distributions align with expectations, i.e., see Figure~1 of \cite{Pevtsovetal2003ApJ}.

For each observational date the typical solar disk radiative and unsigned magnetic fluxes were also characterized, again with errors propagated as described previously. Note, solar disk radiative and magnetic flux measurements were derived from a region comprising $\approx$\,95\% of the visible disk (i.e., see Figure~\ref{fig:SampleFeature}), and is hereafter are referred to as our FD feature. Additionally we point out, prior to FD magnetic field characterizations, sunspot regions were masked (i.e., only fluxes $\lesssim$\,$|10^3|$\,G were considered) to minimize downward biasing of erroneous feature results in our FD typical magnetic field strengths (e.g., see \citealt{WarrenWinebarger2006ApJ}).

\section{Analysis \& Results}\label{sec:MEC_results}

\subsection{Radiative Versus Magnetic Energy}\label{sec:RadVsMag_TypWay}

\begin{figure*}[!t]
\begin{center}
 \includegraphics[scale=0.225]{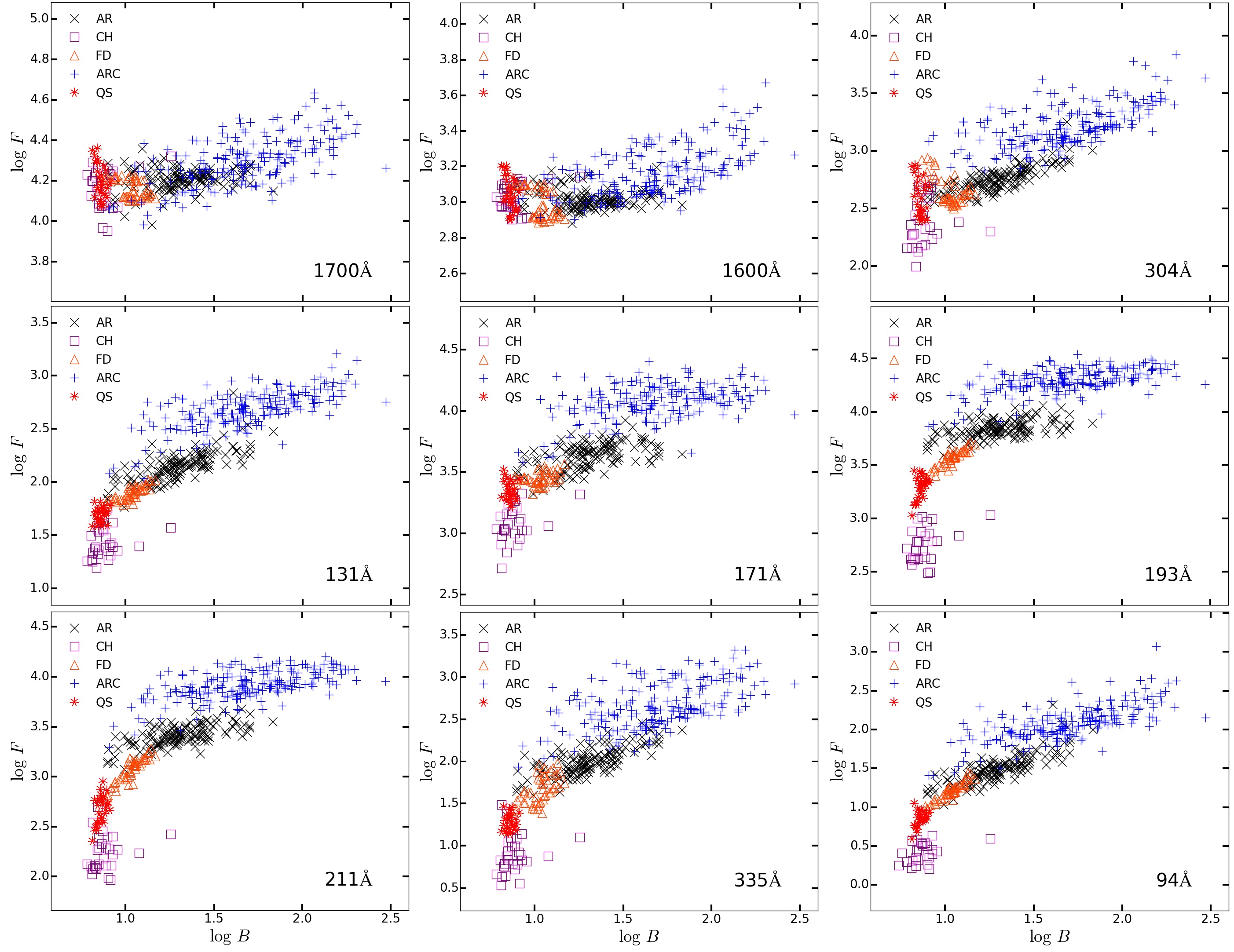}
   \caption{Radiative fluxes (arbitrary units) versus unsigned magnetic flux (arbitrary units) for the 1700\,{\AA}, 1600\,{\AA}, 304\,{\AA}, 131\,{\AA}, 171\,{\AA}, 193\,{\AA}, 211\,{\AA}, 335\,{\AA}, and 94\,{\AA} passbands, from left to right and top to bottom, respectively. On each plot CH, QS, AR, ARC, and FD regions are denoted by {\it squares} (purple), {\it asterisks} (red), {\it x's} (black), {\it pluses} (blue), and {\it triangles} (orange), respectively. Though small spurious feature signal overlap exists, these results emphasize the confidence in our selection methodology of generating statistically significant radiatively differing gross feature samples, while aligning with radiative distribution expectations (e.g., \citealt{Pevtsovetal2003ApJ}). }\label{fig:FvsPhi}
\end{center}
\end{figure*}

In Figure~\ref{fig:FvsPhi} we provide plots of radiative (covering all AIA passbands, with exception of 4500\,{\AA}) versus magnetic fluxes (from HMI observations) with respect to our feature set. Hereafter, we note, the terminologies of chromospheric, TR, and corona are used interchangeably for 304\,{\AA}; 131\,{\AA} and 171\,{\AA}; and 193\,{\AA}, 211\,{\AA}, 335\,{\AA}, and 94\,{\AA} passbands, respectively. Additionally, in terms of 1600\,{\AA}, 1700\,{\AA}, and 4500\,{\AA} observations these are used interchangeably for cooler atmospheric layers (i.e., $\log T$\,$\lesssim$\,4.8; \citealt{Lemenetal2012SolPh}).

It is recognized that AIA passbands are multithermal, and arguments exist for cool and hot emission contamination of a number of passbands (e.g., \citealt{DelZannaetal2011AA,Schmelzetal2013SoPh,Boerneretal2014SoPh}), likely as a function of gross solar atmospheric feature classes (e.g., \citealt{ODwyeretal2010AA}). Important here though, is that predominantly AIA observed radiative fluxes derive from the expected emission line (i.e., \citealt{Lemenetal2012SolPh,Boerneretal2014SoPh}). However, more careful consideration must be taken for 94\,{\AA} and 131\,{\AA} results, as spectral models in the 50 -- 150\,{\AA} range still reveal significant amounts of unaccounted emission, and remain an active area of investigation \citep{Boerneretal2014SoPh}. Outside of flaring conditions, cool emission, originating around 1 MK, is likely contaminating 94\,{\AA} observed fluxes (e.g., \citealt{ODwyeretal2010AA}; \citealt{Boerneretal2014SoPh}), while for 131\,{\AA} hot emission contamination (i.e., Fe\,{\sc xxi} flare line formed around $\log T$\,$\approx$\,7.0) dominates in flaring conditions (e.g., \citealt{DelZannaetal2011AA,DelZanna2013AA,Boerneretal2014SoPh}). In that respect, to first order approximations, that beyond the 94\,{\AA} and 131\,{\AA} observations, AIA passband observations can be assumed to originate from the expected dominant emission lines.

Though not shown, 4500\,{\AA} radiative to magnetic field comparisons provide no evidence of a thermal to magnetic coupling: that is, little to negligible variations in its radiative energy occurs for increasing magnetic field strengths, independent of feature. These results are consistent with the expected high $\beta$ (i.e., ratio of gas to magnetic pressure) conditions that should dominate here.

The results shown in Figure~\ref{fig:FvsPhi} reveal as a function of analyzed features and for solar atmospheric temperatures of $\log T$\,$\lesssim$\,4.8 (i.e., 1700\,{\AA} and 1600\,{\AA} plots therein), minimal radiative energy distinctions\ exist. However, there is a slight ``knee," at approximately $\log B$\,$\sim$\,1.0 where a blending of the radiative energies observed in CH, QS, AR, ARC (to a lesser degree), and FD occurs. We point out that such results are expected, again considering the $\beta$\,$\gtrsim$\,1 conditions that should prevail here \citep{Abbett2007ApJ}. In contrast, the ARC results of these regimes are trending towards a possible linear thermal to magnetic energy relationship (e.g., \citealt{Pevtsovetal2003ApJ}). Thus, relative to other gross features, enhanced ARC photospheric magnetic field strengths could be leading to frozen-in-flux conditions (i.e., $\beta$\,$<$\,1) at cooler atmospheric layers/heights.

In the chromosphere results are consistent with the expectations of a linear radiative to magnetic energy trend, which scales across the gross feature classes. These results reveal an emerging distinction between observed radiances of CH and QS conditions; correlating with similar strengths in their underlying magnetic field energies. Note that the knee identified in cooler passbands remains distinctly discernable in chromospheric emission. For ARs and ARCs, similar observations to the cooler atmospheric layers prevail, i.e., radiative to magnetic flux distributions provide evidence to a linear linking.

In passbands dominated by emission from TR temperatures, with the possibility of lower and/or upper coronal contributions (i.e., 131\,{\AA}; \citealt{ODwyeretal2010AA}; \citealt{DelZannaetal2011AA}; \citealt{Schmelzetal2013SoPh}; \citealt{Boerneretal2014SoPh}), the knee structure of cooler atmospheric layers has ``smoothed" out. However TR radiative versus magnetic energy distributions give rise to signatures of an ankle and knee. The ankle corresponds to CH conditions, which is distributed downward to lower radiative energies than other studied feature classes. This observation is consistent with the work of \citet{Pevtsovetal2003ApJ} for regions dominated by single polarity magnetic fluxes (i.e., see their Figure~1). Note, 131\,{\AA} and 171\,{\AA} provide evidence for similarly distributed CH and QS radiative flux distributions, relative to their respective underlying magnetic field strengths and other studied features. Thereby, elevating arguments adopted here that 131\,{\AA} emission predominantly reflects upper TR regimes in nonflaring conditions, as expected (e.g., \citealt{Lemenetal2012SolPh}). The upper TR knee is emerging where a portion of ARC observations have ``migrated" to higher energies, compared to their AR counterparts. Here we point out, as observed in Figure~\ref{fig:FvsPhi}, 131\,{\AA} to 171\,{\AA}, comparisons particularly of AR and ARC distributions, reveal subtle differences that support hot emission contamination of the 131\,{\AA} passband. Specifically, 131\,{\AA} AR and ARC radiative distributions are more reminiscent to the 211\,{\AA} and 335\,{\AA} passbands.

In the warm corona, described here by AIA's 193\,{\AA} and 211\,{\AA} passbands \citep{Lemenetal2012SolPh,Boerneretal2014SoPh}), results are generally similar to those of the TR. The only distinction of coronal to TR observations exists in a comparison of their CH and QS radiative energy distributions. Both are characterized by decreased radiative energy distributions relative to other analyzed features. The ``upper TR -- coronal" ARC knee suggest the possibility of heating not directly attributable to the magnetic field. This idea aligns with recent challenges to the standard coronal heating model interpretation \citep{Parker1983GApFD}, particularly observations of inversely proportional emission measure (EM) to underlying field strengths of ARCs \citep{Warrenetal2012ApJ}, as well as the absence of non-thermal velocity to temperature trends \citep{BrooksWarren2016ApJ}. Moreover, they provide observational evidence in favor of the presence of unresolved emission (e.g., \citealt{DelZannaMason2003AA}; \citealt{ViallKlimchuk2012ApJ}; {\bf \citealt{Subramanianetal2014ApJ}}).

At hotter coronal temperatures, i.e., those of 335\,{\AA} and 94\,{\AA} observations, with the possibility of cooler emission contributions (i.e., 94\,{\AA}; \citealt{ODwyeretal2010AA,Boerneretal2014SoPh}), similar characteristics prevail to that of the middle corona, and to a lesser degree the TR. In opposition to cooler regions, however, at upper coronal regimes the ARC knee appears smoother. As such a more distinct linear relation of QS, FD, AR, and ARC features, listed in accordance with increasing magnetic field strengths of the respective distributions, is witnessed. We again emphasize; these results, and the previously described upper TR -- coronal ARC trend, further elevate arguments for the existence of unresolved coronal emission (e.g., \citealt{DelZannaMason2003AA}; \citealt{ViallKlimchuk2012ApJ}; {\bf \citealt{Subramanianetal2014ApJ}}). Recognizing the likelihood of cool emission contamination of 94\,{\AA} observations (e.g., \citealt{Boerneretal2014SoPh}), the following artifacts are highlighted. Its CH, QS, and FD radiative distributions qualitatively favor such, based on feature radiative flux distribution consistencies with cooler passbands (Figure~\ref{fig:FvsPhi}). In contrast, however, if such notions held for ARs and ARCs, their respective distributions would be expected to exhibit similarities of cooler atmospheric layers, i.e., the upper TR -- coronal ARC knee, which is clearly not as distinct as for example in 171\,{\AA} or 193\,{\AA} observations.

\begin{figure*}[!t]
\begin{center}
 \includegraphics[scale=0.18]{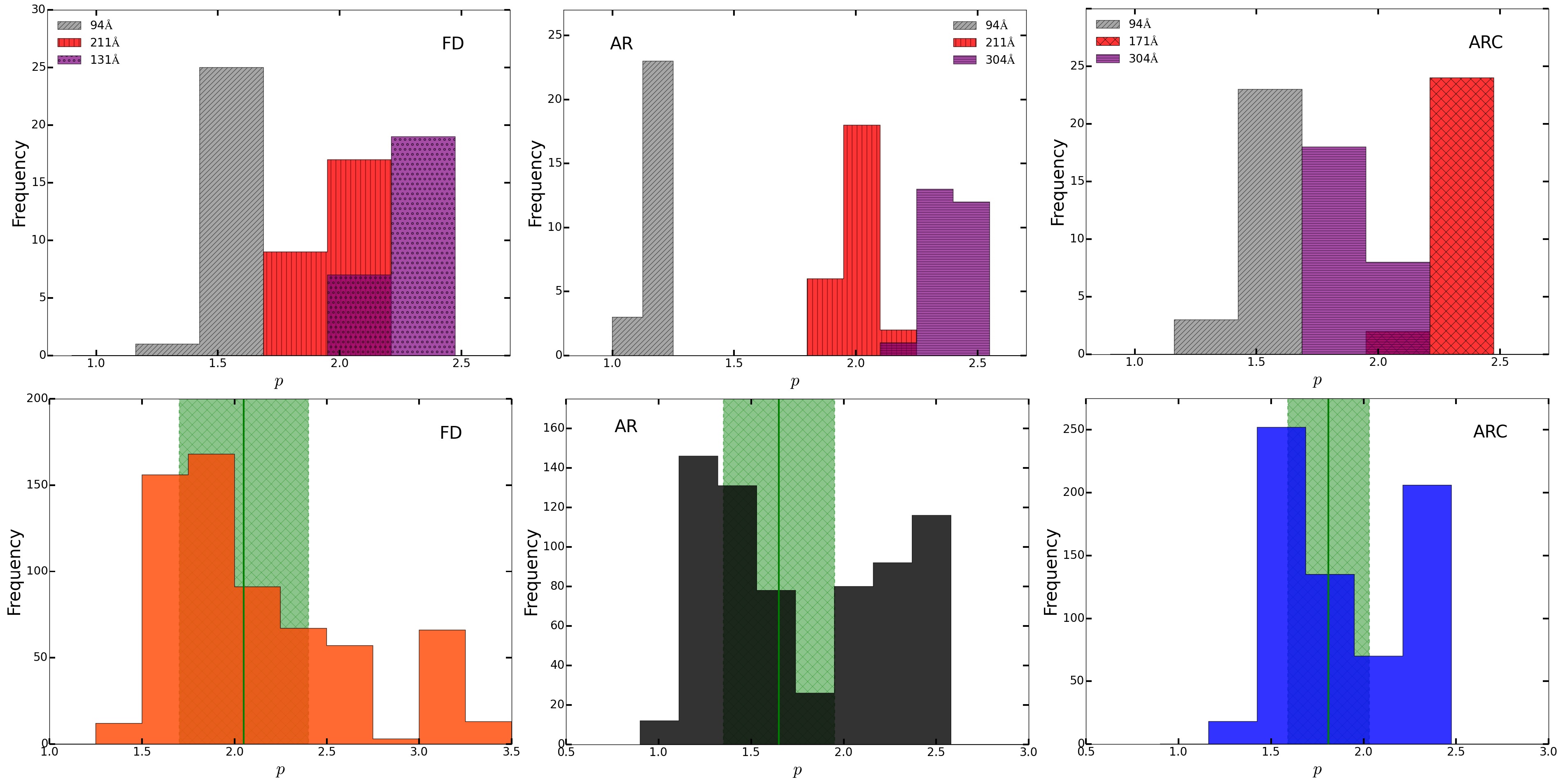}
   \caption{Histogram distributions of Equation~\ref{fig:FvsPhi} derived power-law indices for sample features and passbands. Passband coefficient ensembles (top row) reflect varying physical constraint applications (see $\S$~\ref{sec:RadVsMag_LinearPLaw}), while feature coupling coefficients (bottom row) reflect all analyzed passbands and subset sampling. On the latter panels, the solid (green) vertical lines indicate the features' derived magnetic energy coupling coefficient (Table~\ref{tbl:PowerLawIndicies}), with uncertainties denoted by shaded (green) regions.}\label{fig:FvsPhi_FitExamples}
\end{center}
\end{figure*}

A direct comparison of Figure~\ref{fig:FvsPhi}'s 94\,{\AA} results to Figure 5b of \citet{Benevolenskayaetal2002ApJ} reveals distinctive similarities, particularly our results align with the suggestions of \citet{Benevolenskayaetal2002ApJ} and \citet{FludraIreland2003AA} for two differing dependencies of radiative energy versus that of the underlying magnetic field. As one progresses to cooler atmospheric layers, i.e., the TR to the chromosphere (171\,{\AA} and 304\,{\AA}, respectively) though, our results are reminiscent of the notion of a linear linking of the corona to magnetic fields (i.e., \citealt{Pevtsovetal2003ApJ}). In summary,  radiative versus magnetic flux presentations (Figure~\ref{fig:FvsPhi}), with coverage of the solar atmosphere's gross features and FD scales across broad spectrum regimes, qualitatively favor an extension of linear magnetic coupling, although with the possibility of additional non magnetic heating as revealed by low corona observations.

\subsection{Magnetic Energy Redistribution}\label{sec:RadVsMag_LinearPLaw}

In this section we investigate the magnetic to radiative energy coupling of our gross feature classes and AIA passband observations. We employ the assumption that observed radiative energy results from the dissipation of magnetic free energy (e.g., \citealt{Parker1963ApJS}), and therefore, per $x$, i.e., $x \in \{{\rm CH, QS, FD, AR, ARC} \}$, and AIA passband, $\lambda$, the following linear equation
\begin{equation}\label{eqn:PevFit}
F_{x, \lambda} \propto B^{p_{x,\lambda}},
\end{equation}
was fitted to our data to obtain the energetic magnetic to radiative coupling descriptions, i.e., $p_{x,\lambda}$, in a similar fashion to the previous works of \cite{Golubetal1980ApJ,Hara1996PhDT,Fisheretal1998ApJ,Roaldetal2000ApJ,Wolfsonetal2000ApJ,Schrijver2001ApJ,Benevolenskayaetal2002ApJ,Pevtsovetal2003ApJ,Saar1996IAUS,Vidottoetal2014MNRAS,Alvarado-Gomez2016AA}. Feature coefficients (i.e., $p_x$) were derived from linear least-square regression fits ({\sf MPFIT}; \citealt{Markwardt2009ASPC}) to Equation~\ref{eqn:PevFit} for varying physical constraints (detailed below) as a function of $\lambda$ (i.e., passband; Figure~\ref{fig:FvsPhi_FitExamples}), and then, computing $\langle p_x \rangle$ (Figure~\ref{fig:FvsPhi_FitExamples}) using the bootstrap procedure \citep{Pressetal2002Book}. Therefore, $\langle p_x \rangle$ coefficients were derived similarly to the methodologies of \cite{Pevtsovetal2003ApJ}, as they reflect defining power law indices via least-squares and bootstrap procedures, with steps implemented to provide more realistic coefficients and errors. In relation to latter, this is, they include weighting from statistical and nonstatistical uncertainties from varying physical constraint applications.

The differing physical constraints from which least-squares regression fits were applied, included the following data perturbations: radiative and magnetic flux modulations ($\pm$\,$\leq$\,20\,\%); truncation to upper and lower feature energetic distributions ($\leq$\,10\,\% of observed data); and random observational data subset sampling. Note that our $\langle p_x \rangle$'s reflect radiative and magnetic energy truncations, perturbations which aided in simulating more fully sampled distributions (Figure~\ref{fig:FvsPhi}), their derived power-law slopes include weighting from energetic boundaries expected in the solar atmosphere. Passband magnetic coupling uncertainties per feature (i.e., $p_{x,\lambda}$) were defined as the summation of their fit 1$\sigma$ coefficient deviations and resultant fit errors (Figure~\ref{fig:FvsPhi_FitExamples}). In terms of $\langle p_x \rangle$ uncertainties, they were propagated during passband smoothing via summing the afore described fit ensemble deviations with said subsets standard error on the mean (Figure~\ref{fig:FvsPhi_FitExamples}).

To summarize, our reported coefficients and uncertainties include weighting from nonstatistical errors such as instrument sensitivity, varying physical plasma conditions, and selection bias. We also emphasize that previous works (e.g., \citealt{Pevtsovetal2003ApJ}; \citealt{WarrenWinebarger2006ApJ}) have highlighted the importance of two-side significance from zero measurements as a determination of the fit qualities over that of the $\chi^2$ distribution, in radiative to magnetic energy coupling investigations. Deviations from zero of our measured two-sided significance between radiative and magnetic fluxes from Spearman's rank correlation coefficients for all our power-law indices were statistically significant (i.e., $s$\,$\approx$\,0).

Table~\ref{tbl:PowerLawIndicies} gives our coefficients, as well as literature counterparts, $p^l_{x}$, and range (i.e., $p^l_{x,min}$\,/\,$p^l_{x,max}$). Significant variations of $\langle p_{x} \rangle$ exist between the feature coefficients, as expected, i.e., similar trends prevail for $p^l$ subsets. In contrast to literature, this work embodies $\langle p_{x} \rangle$ results derived from broad electromagnetic spectrum regimes of the solar atmosphere (Figures~\ref{fig:FvsPhi} and \ref{fig:FvsPhi_FitExamples}). Note, reported coefficient and uncertainties, the latter of which is significantly larger than typical works (e.g., \citealt{Fisheretal1998ApJ}; \citealt{Wolfsonetal2000ApJ}; \citealt{Benevolenskayaetal2002ApJ}; \citealt{FludraIreland2003AA}; \citealt{WarrenWinebarger2006ApJ}; \citealt{Warrenetal2012ApJ}), are remarkably consistent with the $p_{\rm real}$ data set of \citet{Pevtsovetal2003ApJ}. We emphasize here that this data set was estimated via object averaging of soft X-ray power-law indices under varying physical constraints, in order to provide more realistic errors.

\begin{table*}[!t]
\begin{center}
\caption{Per feature ($x$) studied in this work, we present power-law indices derived from: all AIA passbands, $\langle p_x \rangle$; AIA's 94\,{\AA} and 335\,{\AA} passbands, $\langle p_{94{\AA}, {\rm x}} \rangle$ and $\langle p_{335{\AA}, {\rm x}} \rangle$, respectively; literature mining, $p^{l}_{x}$; and the subsequent literature reported range (i.e., Min/Max). Note, reported power-law indices have been derived from fits to Equation~\ref{eqn:PevFit}, and are immediately followed by their 1$\sigma$ deviations.} \label{tbl:PowerLawIndicies}
\vspace{0.5mm}
\begin{tabular*}{\textwidth}{@{\extracolsep{\fill}} lcccccc}
\hline
\hline \\[0.25ex]
Feature ($x$) & $\langle p_{{\rm x}} \rangle$ & $\langle p_{94{\AA}, {\rm x}} \rangle$ & $\langle p_{335{\AA}, {\rm x}} \rangle$ & $p^{l}_{x}$ & $p^{l}$ -- Min/Max Range \\ [1ex] 
\hline\\[0.25ex]
CH  & 2.38\,$\pm$\,0.23 & 2.01\,$\pm$\,0.12 & 2.26\,$\pm$\,0.13 & 2.06\,$\pm$\,0.07$^{\dagger_1}$ &           \\[0.75ex]

QS  & 1.73\,$\pm$\,0.73 & 1.69\,$\pm$\,0.06 & 1.91\,$\pm$\,0.12 & 1.74\,$\pm$\,0.21$^{\dagger_2}$ & 0.93/2.03 \\[0.75ex]

AR  & 1.65\,$\pm$\,0.30 & 1.15\,$\pm$\,0.06 & 1.47\,$\pm$\,0.06 & 1.43\,$\pm$\,0.40$^{\dagger_3}$ & 0.98/2.30 \\[0.75ex]

ARC & 1.81\,$\pm$\,0.22 & 1.45\,$\pm$\,0.09 & 1.65\,$\pm$\,0.10 &        ---        &     ---    \\[0.75ex]

FD  & 2.05\,$\pm$\,0.35 & 1.52\,$\pm$\,0.05 & 1.86\,$\pm$\,0.08 & 1.73\,$\pm$\,0.19$^{\dagger_4}$ & 1.47/2.10 \\[0.75ex]

\hline
\end{tabular*}
\end{center}
    $^{\dagger_1}$\citet{Pevtsovetal2003ApJ}\\
    $^{\dagger_2}$\citet{Roaldetal2000ApJ}; \citet{Benevolenskayaetal2002ApJ}; \citet{Pevtsovetal2003ApJ}\\
    $^{\dagger_3}$\citet{Fisheretal1998ApJ}; \citet{Pevtsovetal2003ApJ}; \citet{FludraIreland2003AA}; \citet{WarrenWinebarger2006ApJ}; \citet{Warrenetal2012ApJ}\\
    $^{\dagger_4}$\citet{Wolfsonetal2000ApJ}; \citet{Pevtsovetal2003ApJ}
\vspace{0.05in}
\end{table*}

{\bf We have included 94\,{\AA} and 335\,{\AA} results in Table~\ref{tbl:PowerLawIndicies}, as these passbands' electromagnetic spectrum coverage most closely resemble that detailed by existing works (i.e., X-ray). First, note that a comparison of their results to literature reveal a feature independent alignment, particularly as a ``splitting" of the literature mined counterparts. Thereby, largely, energetic coupling coefficient similarities of this work to existing studies are confined to self-similar analyzed spectrums (Figure~\ref{fig:FvsPhi_FitExamples}).} These results importantly elevate support of our previous speculations (i.e., $\S$~\ref{sec:RadVsMag_TypWay}) for the possibility of wavelength dependence with magnetic energy coupling and/or plasma heating beyond that available from the photospheric magnetic field.

From a qualitative standpoint, we previously highlighted that 94\,{\AA} AR and ARC radiative distributions favored the expected upper coronal origin (i.e., $\approx$\,3 MK; \citealt{Boerneretal2012SoPh}), notions self-consistent with their derived magnetic coupling coefficients (Table~\ref{tbl:PowerLawIndicies}). Specifically, its AR result is remarkably similar to the 1.19 of \cite{Fisheretal1998ApJ}, derived from average Soft X-Ray Telescope (SXT), on {\it Yohkoh} \cite{Tsunetaetal1991SoPh}, radiative fluxes of 333 ARs. {\bf Note that the 335\,{\AA} AR power-law index more appropriately resembles $p^l_{AR}$, considered here to stem from the fact this passband most closely resembles SXT observations. The 94\,{\AA} and 335\,{\AA} ARC coefficients align well with the 1.6 value reported by \cite{WarrenWinebarger2006ApJ} in a soft X-ray AR study, where it was noted that their integrated radiative fluxes were dominated by the brightest AR regions. Given this ARC over AR coefficient to $p^l_{AR}$ consistency, we speculate this results from the fact that the bulk of previous works utilized flux integrations, which possibly biased their analyses towards the AR's most luminous component -- the core. Note that we therefore consider } our analytic approach of investigating the typical 94\,{\AA} in ARs and ARCs predominantly reflects the passbands expected dominance by the hot Fe\,{\sc xx} and {\sc xxiii} emission lines (i.e., $\log T$\,$\approx$\,7.0; \citealt{Boerneretal2012SoPh}).

For $\langle p_{x} \rangle$ results (note including errors) no reported values are grossly disproportionate to expectations, particularly, in consideration of the wide literature ranges (e.g., column 5 of Table~\ref{tbl:PowerLawIndicies}). Akin to said wide literature variances, similar artifacts prevail in our feature coefficient fit ensembles, mainly large uncertainties and wide coefficient ensemble distributions in the presence of broad electromagnetic spectrum regimes (i.e., Figure~\ref{fig:FvsPhi_FitExamples}). Therefore, we find further evidence supporting speculations that radiative to magnetic energy coupling descriptions possibly exhibit a wavelength dependence, particularly as such aligns with the recent numerical simulations of Equation~\ref{eqn:PevFit} to cool main sequence stars by \citet{Alvarado-Gomez2016AA}, for EUV, soft X-ray, and X-ray portions of the spectrum. However, as additionally highlighted above, it can not be ruled out such observations could be suggestive of the presence of an additional plasma heating component, possibly married to the dominant mechanism (e.g., \citealt{Tan2014ApJ,UritskyDavila2014ApJ}).

Similar speculations on the presence of plasma heating beyond the standard flare model in magnetic coupling descriptions were presented by \cite{Pevtsovetal2003ApJ} for their FD indices. They attributed a FD knee and higher coupling coefficient (i.e., $\approx$\,2.06) to CH open field contributions, a value emphasized as remarkably consistent with our $\langle p_{FD} \rangle$\,$\approx$\,2.05. Then, of interest is that our FD radiative fluxes were derived in a manner that sought to minimize sunspots, and thus large open field contributions (e.g., \citealt{WarrenWinebarger2006ApJ}). The 94\,{\AA} FD coefficient of Table~\ref{tbl:PowerLawIndicies} aligns with that reported by \cite{Pevtsovetal2003ApJ}, when they accounted for large-scale open contributions, i.e., 1.61. Coupling these observations with emerging evidence that cooler atmospheric heights are the origin of large scale open field structures \citep{Cranmer2012SSRv}, as well as that 94\,{\AA} observations should predominantly reflect cool emission contamination (i.e., CHs and QS; \citealt{ODwyeretal2010AA}), we emphasize the following. Though our radiative and magnetic energy coupling study supports an extension of the common solar atmospheric magnetic coupling description (i.e., Equation~\ref{eqn:PevFit}; \citealt{Golubetal1980ApJ,Hara1996PhDT,Fisheretal1998ApJ,Roaldetal2000ApJ,Wolfsonetal2000ApJ,Schrijver2001ApJ,Benevolenskayaetal2002ApJ,Pevtsovetal2003ApJ,Saar1996IAUS,Vidottoetal2014MNRAS,Alvarado-Gomez2016AA}) across broad electromagnetic spectrum regimes, it has elevated the necessity of such an investigation seeking to elucidate the possibility of additional plasma heating components and/or a potential spectrum dependence.

\subsection{Magnetic Energy Redistribution -- Revisited}\label{sec:RadVsMag_BrokePLaw}

Here we investigate an extension of the ``universal" X-ray luminosity to unsigned magnetic flux description reported by \citet{Pevtsovetal2003ApJ} (i.e., see their Figure~1), in light of our previous results which suggested the presence of an additional plasma heating component and/or a spectrum dependence (i.e., $\S$~\ref{sec:RadVsMag_TypWay} and \ref{sec:RadVsMag_LinearPLaw}). In that respect, we have modified Equation~\ref{eqn:PevFit} as follows,
\begin{equation}\label{eqn:FvsB_BPL}
F_{\lambda} \propto a_{\lambda} + B^{p_{\lambda}}, \hspace{.1in} |B| > {\rm 10\hspace{.05in}G},
\end{equation}
to include an additional free parameter, $a_{\lambda}$, which exhibits no dependence on the photospheric field strength. Thereby, Equation~\ref{eqn:FvsB_BPL} describes the observed radiative energy via the standard assumption of its linear linkage to the dissipation of free magnetic energy (i.e., $B^{p_{\lambda}}$) coupled with an additional non-magnetic plasma heating component with wavelength dependence (i.e., $a_{\lambda}$). Equation~\ref{eqn:FvsB_BPL}'s lower magnetic energy cutoff (i.e., $|B|$\,$>$\,10 G) reduces this analyses to radiative distributions above the CH and QS knee ($\S$~\ref{sec:RadVsMag_TypWay}), a feature particularly prevalent in hotter atmospheric layers (Figure~\ref{fig:FvsPhi}), previously indicated as possible evidence for a differing dependency of radiative versus magnetic energy (\citealt{Benevolenskayaetal2002ApJ,FludraIreland2003AA,Pevtsovetal2003ApJ} and $\S$~\ref{sec:RadVsMag_TypWay}). As observed in Figure~\ref{fig:FvsPhi}, those radiative distributions above said magnetic energy cutoff favor a predominantly self-similar radiative to magnetic energy coupling (e.g., \citealt{Benevolenskayaetal2002ApJ,Pevtsovetal2003ApJ}), and therefore, derivation of feature dependent magnetic energy coupling coefficients are dropped. Below, further reasons are provided for invoking a lower magnetic energy cutoff.
\begin{figure}[!t]
 \includegraphics[scale=0.25]{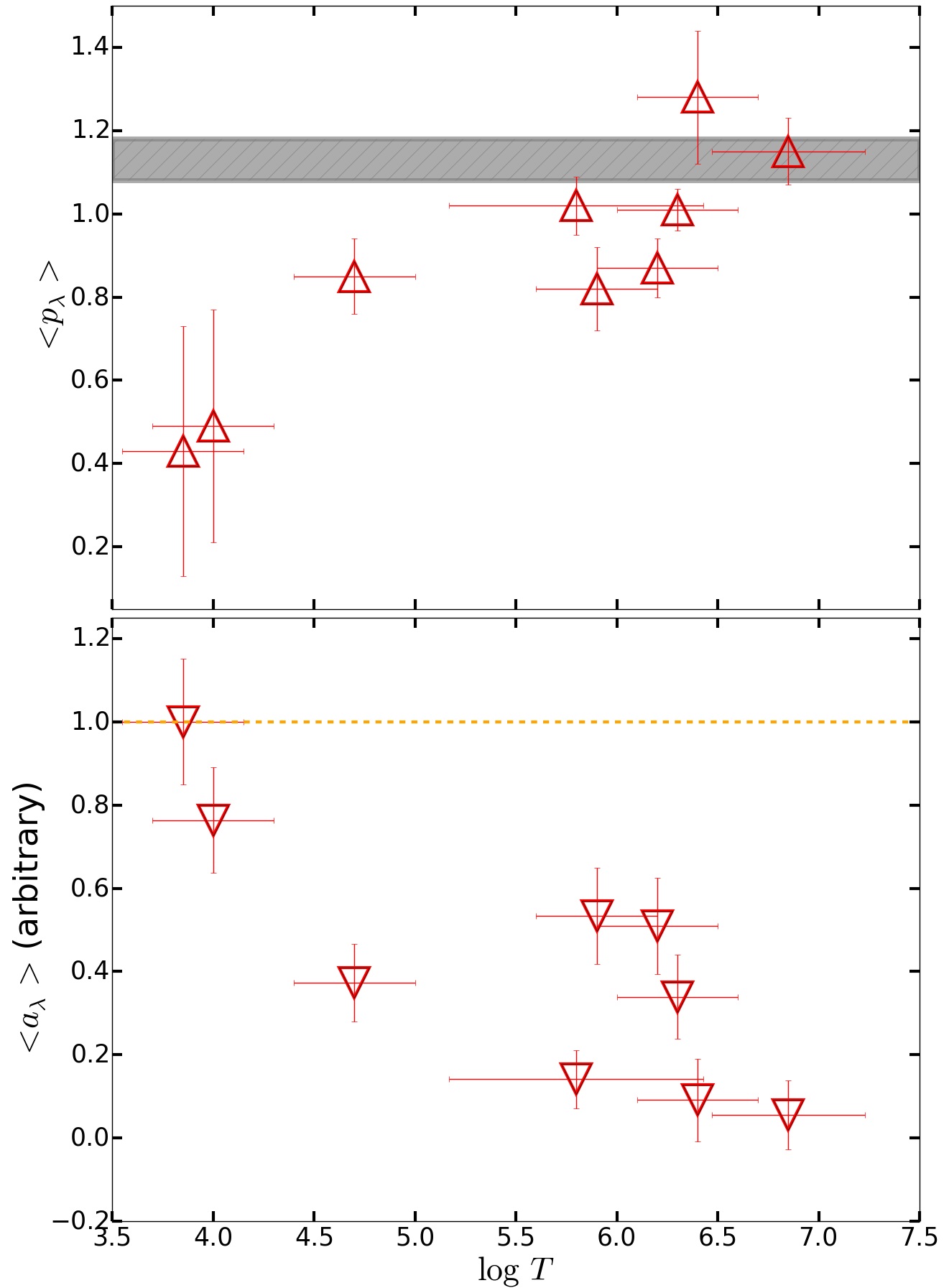}
   \caption{Resultant fit parameters $\langle p_{\lambda} \rangle$ and $\langle a_{\lambda} \rangle$, top and bottom panels, respectively, derived from application of Equation~2 as discussed in the text (see $\S$~\ref{sec:RadVsMag_LinearPLaw}), both plotted as a function temperature for the expected passband dominant emission line (e.g., \citealt{Boerneretal2014SoPh}). Note, in relation to such, passband temperature uncertainties have been assumed at their expected moderate resolution of 0.3 in $\log T$ \citep{Guennouetal2012ApJS}. The exception is 94\,{\AA} and 131\,{\AA}, which both have temperature errors appropriate to their expected cool and hot emission contamination (e.g., \citealt{ODwyeretal2010AA,Boerneretal2014SoPh}), discussed at length in $\S$~\ref{sec:RadVsMag_BrokePLaw}. The shaded gray region on the $\langle p_{\lambda} \rangle$ plot, corresponds to the ``universal" power-law index reported by Pevtsov et al. (2003), i.e., $p$\,$=$\,1.13\,$\pm$\,0.05. Note, we have propagated this universal trend across the entire plotted temperature space, but it was derived from only X-ray observations. To that effect, it potentially highlights a source of previously reported variability in linear magnetic to radiative coupling studies as spectrum dependence.}\label{fig:PLI_SF_VsLogT}
\end{figure}

First, the 10\,G magnetic energy cutoff reduces undesired mathematical applications to our data, i.e., largely avoids diagnostics on clustered data subsets. It aids in reducing nonstatistical effects from instrumental noise, i.e., avoids influences from LOS magnetic fluxes typical of the noise level ($\approx$\,10\,G; A. Sterling 2015; private communication). Note, as observed in Figure~\ref{fig:FvsPhi} below this magnetic, and subsequent radiative boundary our data is a truncation of a hypothetically complete sample. Therefore, theoretically enforcing this lower limit criteria acts to reduce our observations to a more hypothetically complete sample. An upper limit energetic stipulation has been avoided, given previous observationally and numerically centered works (e.g., \citealt{Schrijveretal1989ApJ}; \citealt{Pevtsovetal2003ApJ}; \citealt{Vidottoetal2014MNRAS}; \citealt{Alvarado-Gomez2016AA}) have established the validity of linear magnetic energy coupling to other distant stellar sources, i.e., more radiatively and magnetically energetic.

Equation~\ref{eqn:FvsB_BPL} coefficients were derived similarly to the prescriptions presented in $\S$~\ref{sec:RadVsMag_LinearPLaw}. First, passband coefficient ensembles were obtained from application of least-square fits across the same varying physical constraints, per feature (i.e., FD, AR, and ARC). From these ensembles $\langle a_{\lambda} \rangle$ and $\langle p_{\lambda} \rangle$ were then defined by averaging across various feature subsets (i.e., bootstrapping). Again we point out, as such they reflect results derived from the same object computed under various constraints, and thus provide more realistic assessments as they have been weighted by statistical and nonstatistical effects. Uncertainties in $\langle a_{\lambda} \rangle$ and $\langle p_{\lambda} \rangle$ were propagated as the sum of fit ensemble deviations with the standard error on the mean during subset averaging, and deviations from zero of two-sided significance for all passband to feature sub-samples were statistically significant (i.e., $s$\,$\approx$\,0).

For a direct literature comparison, $\langle p_{\lambda} \rangle$ results were smoothed across various passband subsets, and are summarized as follows. For 94\,{\AA} and 335\,{\AA} passbands, a magnetic energy coupling coefficient of $\langle p \rangle$\,$=$\,1.21\,$\pm$\,0.17 was found, a result consistent with: the 1.19 reported by \cite{Fisheretal1998ApJ} in an X-ray study of 333 ARs; the 1.29 of \cite{Alvarado-Gomez2016AA} obtained from soft X-ray simulations of cool stars (i.e., 2 -- 30\,{\AA}); and the universal description of \cite{Pevtsovetal2003ApJ}, i.e., 1.13. Again, consistent with 94\,{\AA} results presented in $\S$'s~\ref{sec:RadVsMag_TypWay} and \ref{sec:RadVsMag_LinearPLaw} these observations are considered support of previous arguments that the radiative distributions investigated here (i.e., those accompanied by underlying field strengths $>$\, 10 G) predominantly reflect emission from hot coronal temperatures (i.e. $\approx$\,3 MK). For AIA passbands typical of cooler atmospheric layers (i.e., 304, 131, 171, 193, and 211\,{\AA}), we determined \,0.91\,$\pm$\,0.16, a result agreing with \citet{Pevtsovetal2003ApJ}'s reports for XBPs, QS (no averaging), and dwarf stars, as well as the X-ray and EUV simulations of \citet{Alvarado-Gomez2016AA}, i.e., 5 -- 100\,{\AA} and 100 -- 920\,{\AA}, respectively. As highlighted in $\S$~\ref{sec:RadVsMag_TypWay}, 131\,{\AA} observations, mainly in relation to the feature radiative distributions focused on in this analysis (i.e., FD, AR, and ARC), favored hot emission contamination (e.g., \citealt{ODwyeretal2010AA,Boerneretal2014SoPh}). However, marginal power-law index variations in the cooler atmospheric group result if 131\,{\AA} is avoided, while weighting the hotter passbands by its coefficients, leads to $\langle p \rangle$\,$\approx$\,1.15. Thereby, as previously speculated, 131\,{\AA} observations favor multi-thermal emission, particularly in relation to radiative fluxes associated with $|B|$\,$>$\,10 G. In terms of AIA passbands reflective of TR and cooler atmospheric regimes, with avoidance of 131\,{\AA}, $\langle p \rangle$\,$\approx$\,0.65. This aligns quite well with the EUV cool star simulation of \citet{Alvarado-Gomez2016AA}. Independent of passband, and avoiding 1600\,{\AA} and 1700\,{\AA}, a power-law index of \,1.0\,$\pm$\,0.2 is measured. This result is emphasized as support for self-similar plasma heating of the predominantly closed field corona (e.g., \citealt{Pevtsovetal2003ApJ}), while elevating evidence for an extension to cooler atmospheric layers.

Figure~\ref{fig:PLI_SF_VsLogT} presents Equation~\ref{eqn:FvsB_BPL} derived coefficients as a function of passband, where passbands have been plotted in accordance with their expected dominant emission lines formation temperature (e.g., \citealt{Boerneretal2014SoPh}), and given enhanced uncertainties reflective of their expected moderate temperature resolution, i.e., 0.3 in $\log T$ \citep{Guennouetal2012ApJS}. Furthermore, in line with our above discussion, as well as with the results of $\S$'s~\ref{sec:RadVsMag_TypWay} and \ref{sec:RadVsMag_LinearPLaw}, within Figure~\ref{fig:PLI_SF_VsLogT} both 94\,{\AA} and 131\,{\AA} temperature uncertainties have been modified to reflect hot and cool emission contamination. Specifically, we have enhanced their temperature errors as the difference described by \cite{ODwyeretal2010AA,ODwyeretal2012AA} to those reported by \cite{Boerneretal2012SoPh}. Distinctly interesting to our $\langle p_\lambda \rangle$ results, when observed across large solar atmospheric spectrum and thus, temperature regimes (Figure~\ref{fig:PLI_SF_VsLogT}), is that it favors a linear correlation, i.e.,
\begin{equation}\label{eqn:P=TG}
\langle p \rangle \propto T^\gamma,
\end{equation}
where $\gamma$ would be a proxy for the efficiency of magnetic energy redistribution with temperature. It is emphasized that such a functional dependence of the efficiency of magnetic energy deposition with thermodynamic conditions (Figure~\ref{fig:PLI_SF_VsLogT}) are results previously speculated on, e.g., \citealt{Longcope1998ApJ}; \citealt{LongcopeKankelborg1999ApJ}. Then, along with the fact that large statistical samples were utilized in deriving said energetic coupling descriptions (e.g., \citealt{Rosneretal1978ApJ,Dere1982SoPh}), we find confidence in Equation~\ref{eqn:P=TG}'s validity. Even in the presence of the moderate temperature resolution provided by the bulk of AIA channels, and the likelihood of significant cool/hot emission contamination from the 94\,{\AA} and 131\,{\AA} passbands. Further contributing to these speculations is the upper TR -- low coronal dip, ``ankle," witnessed in Figure~\ref{fig:PLI_SF_VsLogT}'s $\langle p_\lambda \rangle$ panel. Specifically, said feature is possibly reminiscent to the: expected upper TR peak in current dissipation per particle \citep{Hansteenetal2010ApJ718,BingertPeter2011AA}; switch in EM to underlying magnetic field strength dependence \citep{Warrenetal2012ApJ}; and/or enhanced coronal abundances resultant from atmospheric energy redistribution processes (e.g., \citealt{Bradshaw2003PhDT,Laming2004ApJ}). Possibly more important to the ankle is its remarkable alignment with the atmospheric temperature regime where significant unresolved emission is expected to reside (e.g., \citealt{DelZannaMason2003AA,ViallKlimchuk2012ApJ}; {\bf \citealt{Subramanianetal2014ApJ}}); discussions deferred to the proceeding section. As the overarching goal of this work remains an investigation of magnetic energy coupling across broad electromagnetic spectrum regimes, we applied no methodologies seeking explicit estimates of temperature to derived energy coupling coefficients. We emphasize such  analyses would introduce further nonstatistical biases given their reliance on systematic calibration assumptions for narrowband observations. Additionally in support of these arguments, spectral model wavelength ranges correlating with AIA channels where the bulk of our multi-thermal emission contamination is expected to reside, i.e., 50 -- 170\,{\AA} (CHIANTI; \citealt{Dereetal1997AAPS}), remain deficient \citep{Boerneretal2014SoPh}.

In terms of the derived $\langle a_{\lambda} \rangle$ coefficients (Figure~\ref{fig:PLI_SF_VsLogT}), we first note its hot corona results (i.e., 94\,{\AA} and 335\,{\AA}) possibly explains well that this work's Equation~\ref{eqn:PevFit} derived coefficient similarities with literature were predominantly constrained to self-consistent spectrum regimes, i.e., Figure~\ref{fig:PLI_SF_VsLogT} favors a less significant role of $a_{\lambda}$ in these passbands. For cooler atmospheric layers, mainly where $\beta$\,$\gtrsim$\,1 conditions should prevail (i.e., 1600\,{\AA} and 1700\,{\AA}), $\langle a_{\lambda} \rangle$ coefficients support expectations of plasma heating not dominated by the magnetic field, i.e., consistent with their marginal magnetic energy coupling coefficients compared to other AIA channels. At temperature regimes intermediate to the afore described results, correlating with the highlighted $\langle p \rangle$ ankle, an increased contribution to observed radiative fluxes from $\langle a_{\lambda} \rangle$ occurs. Therefore, indicating as hypothesized in $\S$'s~\ref{sec:RadVsMag_TypWay} and \ref{sec:RadVsMag_LinearPLaw}, a possible rise in plasma heating contributions not directly attributable to the freely available magnetic energy, however, likely intimately linked to it.

\begin{table}[!t]
\begin{center}
\caption{Spearman rank correlation coefficients ($r_s$) and two-sided significance of its deviation from zero ($s$) tabulated between $a_\lambda$ and typical AR, QS, and FD (defined here as average of QS and AR).}
\label{tbl:ALamb-DEMCorr}
\vspace{0.5mm}
\begin{tabular*}{\columnwidth}{@{\extracolsep{\fill}} lcr}
\hline \hline \\ [0.10ex]
Feature & $r_s$ & $s$  \\
\hline \\ [0.10ex]
AR & 0.69 & 0.06 \\
QS & 0.75 & 0.05 \\
AR/QS Average & 0.85 & 0.01 \\
\hline
\end{tabular*}
\end{center}
\end{table}

Here we speculate on the source of an additional plasma heating component, as possibly revealed by the function form of $\langle a_{\lambda} \rangle$. That is, it exhibits: a high index (``energy") like tail, correlating with cooler atmospheric layers; a TR upturn; an approximate upper TR to lower corona peak; and thereafter, decreases for increasingly hotter temperatures (Figure~\ref{fig:PLI_SF_VsLogT}). Interesting to this $a_\lambda$ description is its resemblance to a ``typical" solar atmospheric differential emission measure (DEM) distribution, e.g.,
\begin{equation}\label{eqn:DEM}
DEM(n_e,T)= n_e^2  \frac{dh}{dT},
\end{equation}
with $h$ the LOS coordinate and $n_e$ the electron density (e.g., see \citealt{ODwyeretal2010AA}), which provide significant insight regarding solar atmospheric thermal structuring. Table~\ref{tbl:ALamb-DEMCorr} presents the Spearmans rank correlation coefficients, $r_s$, and two-sided significance of its deviation from zero, between $a_\lambda$ results to solar atmospheric QS and AR DEMs obtained from the CHIANTI atomic database (e.g., \citealt{Dereetal1997AAPS}), as well as an average of those two features. Though Table~\ref{tbl:ALamb-DEMCorr}'s correlation coefficients represent crude approximations, in relation to this study they elevate the possible connectivity of $a_\lambda$ to the solar atmosphere's thermodynamic conditions, based on their strong correlations. This investigation, therefore, possibly highlights an entanglement of thermodynamic and magnetic energy contributions in previous energetic coupling studies (Equation~\ref{eqn:PevFit}), particularly for descriptions of broad plasma conditions (i.e., AR vs QS, etc.) and spectrum regimes (i.e., soft X-ray through UV), as our results (Figure~\ref{fig:PLI_SF_VsLogT}) present reasonably well the expected manifestation of diffuse unorganized emission at coronal temperatures {\bf \citep{Subramanianetal2014ApJ}}.

\begin{figure*}[!t]
\begin{center}
 \includegraphics[scale=0.23]{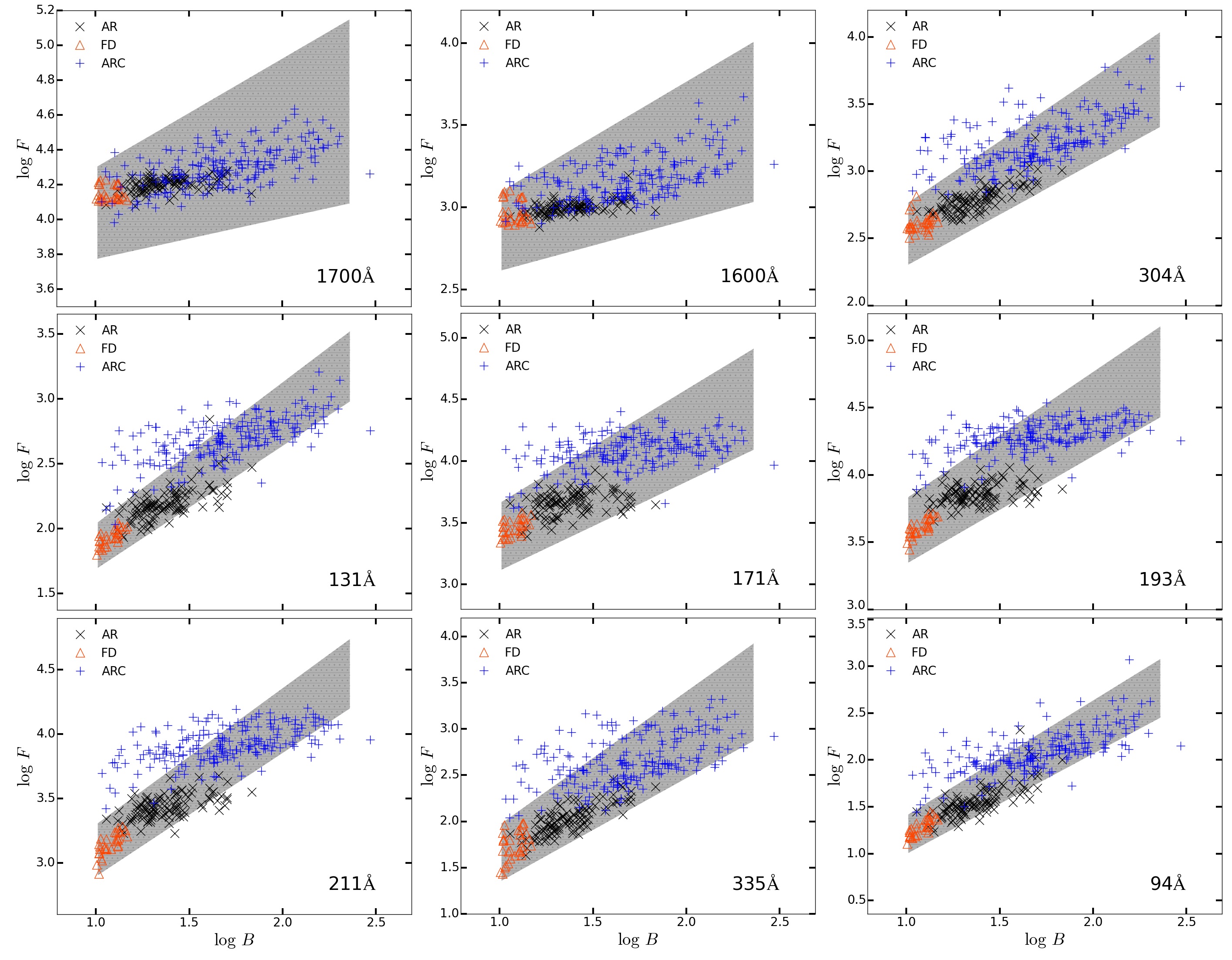}
   \caption{Same as Figure~2, except observational data $<$\,10\,G have been avoided (i.e., $\S$~\ref{sec:RadVsMag_BrokePLaw}. Note, the shaded region represents the fit of Equation~2 to our results, which is extensively discussed in the text, and its respective parameter spaces are presented in Figure~\ref{fig:PLI_SF_VsLogT}.}\label{fig:FvsPhi_2Fits}
\end{center}
\end{figure*}

In Figure~\ref{fig:FvsPhi_2Fits} the fits of Equation~\ref{eqn:FvsB_BPL} to our data, i.e., $\langle a_{\lambda} \rangle$ and $\langle p_{\lambda} \rangle$ per AIA passband, are presented. As observed the fits are consistent with expectations, notions largely confined to X-ray portions of the spectrum (to the best of our knowledge). That is, across broad solar atmospheric electromagnetic spectrum regimes radiative fluxes approximately linearly scale with those of the underlying magnetic field; albeit with varying magnetic energy coupling descriptions as a function of the electromagnetic spectrum (i.e., Figure~\ref{fig:PLI_SF_VsLogT}). These results are akin to the numerically derived results of \cite{Alvarado-Gomez2016AA}, who reported a wavelength dependence in the magnetic energy redistribution of cool main sequence stars, and whose work further elevates arguments presented here for Equation~\ref{eqn:P=TG} (Figure~\ref{fig:PLI_SF_VsLogT}). It is recognized, energetic scatter about Equation~\ref{eqn:FvsB_BPL} descriptions remain in Figure~\ref{fig:FvsPhi_2Fits}, predominantly correlated with ARCs. An artifact of interest as it aligns well with spectrum regimes favored by numerous ARC works as the location of significant unresolved emission (e.g., \citealt{DelZannaMason2003AA}; \citealt{ViallKlimchuk2012ApJ}; {\bf \citealt{Subramanianetal2014ApJ}}; \citealt{Alvarado-Gomez2016AA}).

\subsection{Coronal Heating}\label{sec:CoronalHeatingApp}

First, recall an $\langle p_\lambda \rangle $ ankle was highlighted in $\S$~\ref{sec:RadVsMag_BrokePLaw} that correlated well with spectral and temperature regimes favored as the location of significant unresolved coronal emission (e.g., $\log T$\,$\approx$\,6.0 -- 6.5; \citealt{DelZannaMason2003AA}; \citealt{ViallKlimchuk2012ApJ}; {\bf \citealt{Subramanianetal2014ApJ}}), and aligned with a ``bump" in our $\langle a_\lambda \rangle$ distribution (Figure~\ref{fig:PLI_SF_VsLogT}). Hereafter the feature is referred to as $\log T_w$, and we hypothesize that as Equation~\ref{eqn:P=TG} favors that across the bulk of solar atmospheric temperature space the efficiency of magnetic energy redistribution approximately linearly scales. As the compilation of results presented in this work support, $\log T_w$ is considered to correlate with upper TR -- low corona temperatures. To support such assumptions, we point out our Equation~\ref{eqn:P=TG}'s alignment with previously established linear correlations of temperature to EM distributions (e.g., \citealt{Warrenetal2012ApJ}; {\bf \citealt{Subramanianetal2014ApJ}}; \citealt{DelZannaetal2015AA}), and pressure and loop length (e.g., \citealt{Rosneretal1978ApJ}; \citealt{KanoTsuneta1995ApJ}). Thereby, it is necessary here to explain the ``obscured" $\log T_w$ radiative observations in our magnetic coupling descriptions, where observational evidence indicates that ``cool" plasma excess exists (i.e., Figure~\ref{fig:PLI_SF_VsLogT}). {\bf Below we present a generalized coronal heating theory, that centers on the dominant energy sources, i.e., magnetic, enthalpy, and thermal conduction \citep{BradshawCargill2010ApJ}.}

\begin{figure}[!t]
\includegraphics[scale=0.295]{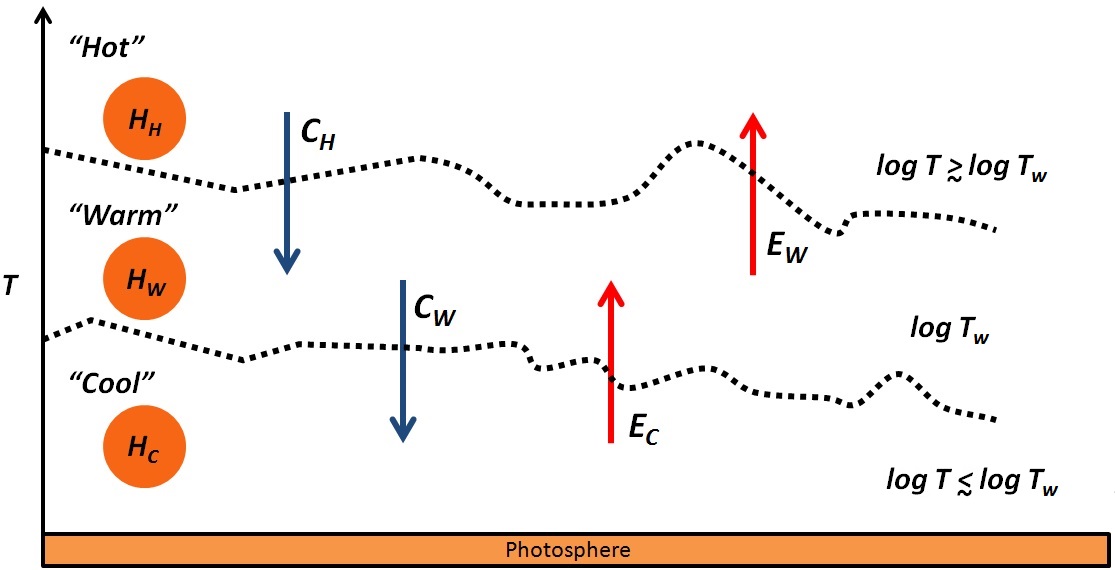}
   \caption{Cartoon schematics of our proposed theoretical solar atmospheric heating descriptions and subsequent energy redistribution processes.}\label{fig:CorHeat_Model}
\end{figure}

Consider a generalized and hypothetical solar atmosphere segmented into cool ($\log T$\,$\lesssim$\,$\log T_w$), warm ($\log T_w$), and hot ($\log T$\,$\gtrsim$\,$\log T_w$) layers (Figure~\ref{fig:CorHeat_Model}), each of which experiences local heating via the freely available magnetic energy (i.e., $H_{cool}$, $H_{warm}$, and $H_{hot}$, respectively; Equation~\ref{eqn:P=TG}). {\bf Within the cool layer, atmospheric heating, via chromospheric evaporation, $E_{cool}$ }(e.g., \citealt{Fisheretal1985ApJ}; \citealt{CraigMcClymont1986ApJ}; \citealt{Hansteenetal2010ApJ718}), would contribute to warm enhanced plasma emission. Additionally, under the standard coronal heating picture (e.g., \citealt{Oluseyietal1999ApJ524-1105O}; \citealt{Oluseyietal1999ApJ527-992O}), downward conducted hot layer heat flux ($C_{hot}$) would provide a source of radiatively bright warm emission. Assuming for simplicity that heated evaporating plasma ($E$) and conduction ($C$) processes represent a portion ($\delta$) of local layer heating, we arrive at a total warm heating ($H^t_{warm}$) given by
\begin{equation}\label{eqn:WarmLayerHeat}
H^t_{warm} \approx H_{warm} + \delta H_{cool} + \delta H_{hot}.
\end{equation}
Using similar arguments the total cool and hot heating would be described as
\begin{equation}\label{eqn:CoolLayerHeat}
H^t_{cool} \approx H_{cool} + \delta H_{warm},
\end{equation}
and
\begin{equation}\label{eqn:HotLayerHeat}
H^t_{hot} \approx H_{hot} + \delta H_{warm},
\end{equation}
respectively (Figure~\ref{fig:CorHeat_Model}). In other words, warm conduction and evaporation contributes to the cool and hot regions, respectively, while local plus hot and cool energy redistribution processes contribute to the warm layer (Figure~\ref{fig:CorHeat_Model}). We find support for these general arguments in works that have favored the presence of unresolved emission, particularly correlating with our proposed $\log T_w$ space (e.g., \citealt{DelZannaMason2003AA}; \citealt{ViallKlimchuk2012ApJ}; \citealt{Warrenetal2012ApJ}; {\bf \citealt{Subramanianetal2014ApJ}}; \citealt{DelZannaetal2015AA}; \citealt{Alvarado-Gomez2016AA}). We speculate that this enhanced volume of warm heated plasma, relative to our other layers (i.e., Figure~\ref{fig:PLI_SF_VsLogT}), leads to a $\log T_w$ magnetic energy redistribution efficiency ankle. More specifically, this is, the $p$ versus $\log T$ efficiency ankle would be expected to manifest as diffuse ``unorganized" emission, i.e., the $a_\lambda$ bump (Figure~\ref{fig:PLI_SF_VsLogT}); notions expounded upon below.

In terms of the predominantly closed field corona, various scales of closed magnetic flux tubes exist, rooted in the network or intranetwork lanes, and extending to various heights ($z$) from the solar photosphere  (e.g., \citealt{Oluseyietal1999ApJ524-1105O}; \citeyear{Oluseyietal1999ApJ527-992O}; \citealt{Orangeetal2010AAS}; \citeyear{Orangeetal2011AAS}; \citealt{Tan2014ApJ}). The classical one-dimension steady state loop energy equation (e.g., \citealt{Rosneretal1978ApJ}; \citealt{CraigMcClymont1986ApJ}) can be written in the following conservative form
\begin{equation}\label{eqn:EngConvs}
H(T) = \frac{d}{dz}\left[ 5 n_e v k_B T + F_c \right] + n^2_e \Lambda(T),
\end{equation}
where $H$ defines the energy input (i.e., herein local heating defined previously), and $n_e$, $v$, $k_B$, $T$, $F_c$, and $\Lambda(T)$ represent the electron density, velocity, Boltzman constant, temperature, conductive flux and radiative loss function, respectively. Note that in Equation~\ref{eqn:EngConvs} the following assumptions have been made. {Loops are assumed to be in hydrostatic equilibrium, such that gravity is balanced by the pressure gradient.}
Flows are subsonic, i.e., $v / c$\,$<$\,1, and low Mach numbers prevail ($M^2$\,$<<$\,1), implying that the kinetic energy density is small compared to the thermal energy density. Finally, we have ignored non-uniformities in loop areas, thus, loop cross section factors have been assumed to be on the order of unity. Equation~\ref{eqn:EngConvs} shows that the energy source ($H$) supports depletion of enthalpy, heat conduction fluxes, and radiative losses.

The classical Spitzer conductivity for full ionized plasmas is appropriate, i.e.,
\begin{equation}\label{eqn:CondFlux}
F_c = - \kappa T^{5/2} \frac{dT}{dz},
\end{equation}
with $\kappa$\,$\sim$\,10$^{-6}$ for $\log T$\,$\ge$\,5.0, while for cooler regimes the effects of ambipolar diffusion on the total particle heat flux should be considered (\citealt{Fontenlaetal1990ApJ}; \citeyear{Fontenlaetal1991ApJ}; \citeyear{Fontenlaetal1993ApJ}). The radiative loss function has been analytically approximated by sequenced power laws of the form
\begin{equation}\label{eqn:RadLoss}
\Lambda (T) = \Lambda_s (T / T_s)^M,
\end{equation}
joined continuously (e.g., see \citealt{Oluseyietal1999ApJ524-1105O}; \citeyear{Oluseyietal1999ApJ527-992O}). Using the continuity equation,\
\begin{equation}\label{eqn:ContinuityEqn}
\frac{d}{dz} \left[ n_e v \right] = 0,
\end{equation}
and simplifying Equation~\ref{eqn:EngConvs}, we arrive at a energy balance form given by
\begin{equation}\label{Eqn:SimplEngCons}
H(T) = 5 k_B q \frac{d T}{dz} + \frac{d F_c}{dz} + n^2_e \Lambda(T),
\end{equation}
with $q$\,$=$\,$n_e v$. Noting the common solar atmospheric temperature stratification, where
\begin{equation}\label{eqn:AtmThermGrads}
\frac{d T}{dz} > 0,
\end{equation}
for increasing $z$ (e.g., \citealt{Murawskietal2013MNRAS}; \citealt{Orange2014PhDT}), leads to the following condition
\begin{equation}\label{eqn:ModelPressGrad}
\left(\frac{dT}{dz}\right)_{cool - warm} >> \left(\frac{dT}{dz}\right)_{hot - warm},
\end{equation}
in relation to our model (Figure~\ref{fig:CorHeat_Model}). In that respect, we consider that conductive ($C$) and evaporative ($E$) processes then most strongly reflect
\begin{equation}
C \propto \frac{d^2}{dz^2}\left[T^{\frac{7}{2}}\right],
\end{equation}
and
\begin{equation}\label{eqn:EvapTodTdz}
E \propto \frac{dT}{dz},
\end{equation}
respectively. Directly then Equations~\ref{eqn:AtmThermGrads} -- \ref{eqn:EvapTodTdz} lead to the following conditions
\begin{equation}
H_{hot} > H_{warm} > H_{cool};
\end{equation}
thus,
\begin{equation}
H_{hot} >> H_{cool},
\end{equation}
while
\begin{equation}
C_{warm} >> C_{hot},
\end{equation}
and
\begin{equation}
E_{cool} >> E_{warm}.
\end{equation}
Equations~\ref{eqn:WarmLayerHeat} -- \ref{eqn:HotLayerHeat} can then be simplified to 
\begin{equation}\label{eqn:SimpHot}
H^t_{hot} \sim H_{hot},
\end{equation}
\begin{equation}\label{eqn:SimpWarm}
H^t_{warm} \sim H_{warm} + E_{cool},
\end{equation}
and
\begin{equation}\label{eqn:SimpCool}
H^t_{cool} \sim C_{warm}.
\end{equation}
We therefore, arrive at three solutions, ``classes," that should dominate observational signatures in light of our proposed model, i.e., hot local, warm local plus chromospheric evaporation, and cool conductive back heating, for decreasing $z$, respectively. These solutions lend further support to speculations on the source of the $\log T_w$ magnetic energy redistribution dip. We also emphasize, their alignment with the proposed three classes of loop solutions presented by \citet{Oluseyietal1999ApJ527-992O}, i.e.,
\begin{enumerate}
\item[1.] Radiation dominated, where $H$ is large and $d T/dz$ is small,
\item[2.] Classical, where both $H$ and $d T/dz$ are intermediate, and
\item[3.] Conduction dominated, where $H$ is small and $d T/dz$ is large,
\end{enumerate}
respectively. It is recognized that no discussions were presented in relation to the driver of our hypothesized local heating ($H$), nor temporal variability, and as such we briefly address these below.

The Sun's atmosphere is not hydrostatic (\citealt{Oluseyietal1999ApJ524-1105O}; \citeyear{Oluseyietal1999ApJ527-992O}), nor does the corona's mass decrease over time \citep{Hansteenetal2010ApJ718,Guerreiro2013ApJ}. Therefore, we would expect that time-dependent heating processes, such as postulated here, would lead to near continuous plasma heating/cooling and mass redistribution \citep{Hansteenetal2010ApJ718}, e.g., plasma heating driven by field line stress build up and dissipation (e.g., \citealt{Klimchuk2006SoPh}) from photospheric convective motions and magnetic field recycling \citep{Berger1997PhDT,Bergeretal1997SPD}. We speculate from a purely theoretical standpoint, as our model provides no insight into temporal variability, that the existence of pervasive TR plasma downflows and marginal coronal upflows (e.g., \citealt{Doscheketal1976ApJ,Dereetal1984ApJ,Achouretal1995ApJ,Hansteenetal1996ASPC,PeterJudge1999ApJ,DelZanna2008AA,Doscheketal2008ApJ,Tripathietal2009ApJ,Hansteenetal2010ApJ718})
should be expected. {\bf In the framework of our model, $\log T_{cool}$, hypothetically represents a region comprised in large by conduction dominated loops, while the classical type would be expected to dominate in $\log T_w$, i.e., a compilation of magnetic to thermal energy conversions and evaporation processes. Recognizing that our model has centered on dominant energy sources, it is also just as probable that wide-spread TR downflows could arise from enthalpy cooling processes \citep{BradshawCargill2005AA,BradshawCargill2010ApJ,BradshawCargill2010ApJ717}.}

Recognizing the presence of knee and ankle structures (Figure~\ref{fig:FvsPhi}), and energetic scatter about Equation~\ref{eqn:FvsB_BPL} descriptions (Figure~\ref{fig:FvsPhi_2Fits}), we would expect additional plasma heating beyond that provided by footpoint motions (Figure~\ref{fig:PLI_SF_VsLogT}). For example, the dominant heating mechanism married to a diffuse background component as proposed in recent works \citep{UritskyDavila2014ApJ,Tan2014ApJ}, and/or plasma heating via magnetohydrodynamic wave dissipation (e.g., \citealt{HollwegYang1988JGR}; \citealt{PoedtsGroof2004ESASP}). Furthermore, such additional plasma heating components could be related to enhanced coronal elemental abundances (e.g., \citealt{Bradshaw2003PhDT}), driven by MHD turbulent dissipation as described by \cite{Laming2004ApJ,Laming2009ApJ,Laming2012ApJ}.

It is noted, the role of radiative losses in our model discussions have been avoided. Radiative loss energy depletion scales as the density squared (Equation~\ref{Eqn:SimplEngCons}), while it is widely known that $n_e$ decreases for increasing $z$ \citep{Abbett2007ApJ}. Thus, cooler loops of our proposed model, characterized by apex densities greater than their hot counterparts \citep{Hansteenetal2014Sci}, would then experience more efficient energy depletion via radiative losses, compared to thermal conduction \citep{Spadaroetal2006ApJ,Hansteenetal2010ApJ718,Guerreiro2013ApJ}. In that respect, for our model, local heating presumably occurs self-consistently across all layers, where progressively cooler atmospheric layers experience more efficient radiative cooling \citep{Abbett2007ApJ,Hansteenetal2014Sci}.

\section{Conclusions \& Summary}\label{sec:MEC_SummaryConclusion}

Observational data from all available AIA passbands and HMI LOS magnetograms at approximately 3 -- 5 day intervals during May 2010 -- July 2013 were utilized to characterize the typical radiative and magnetic fluxes, respectively, of CH, QS, AR, ARCs, and at FD scales. Note that these data provided coverage of visible (i.e., photospheric) through soft X-ray (i.e., upper coronal plasmas) spectrum of the predominantly open and closed coronal magnetic field environments of the gross feature classes dominating the solar disk, independent of activity phase.

Radiative to magnetic energy coupling assessments were carried out: first, per feature (i.e., $\langle p_{x} \rangle$), by deriving power law indices from all AIA passbands, barring 4500\,{\AA}, and smoothing across UV through X-ray observations (i.e., $\S$~\ref{sec:RadVsMag_LinearPLaw}); and secondly, per AIA passband (i.e., $\langle p_{\lambda} \rangle$) from power-law indices averaged across radiative to magnetic energy distribution fits with a 10 G lower magnetic energy cutoff (i.e., $\S$~\ref{sec:RadVsMag_BrokePLaw}). For the first approach, i.e., with feature dependence, results revealed consistencies to existing literature (Table~\ref{tbl:PowerLawIndicies}) at similarly analyzed energy ranges (i.e., soft X-ray; e.g., \citealt{Pevtsovetal2003ApJ}), and highlighted the possibility of a wavelength dependence and/or plasma heating contributions not directly related to the available magnetic field strength ($\S$~\ref{sec:RadVsMag_LinearPLaw}). Application of Equation~\ref{eqn:FvsB_BPL} to a sub-sample of our features, e.g., passband dependent derived power-law indices, $\langle p_\lambda \rangle$, provided reasonable approximations of magnetic energy redistribution across previously unexplored (to best of our knowledge) spectrum regimes (Figure~\ref{fig:FvsPhi_2Fits}), and supported self-similar heating of the large scale closed field corona (e.g., \citealt{Klimchuk2015RSPTA}). However,
our analysis of Equation~\ref{eqn:FvsB_BPL} also favored the presence of an additional passband dependent non-magnetic plasma heating component, i.e., $a_\lambda$ (Figure~\ref{fig:PLI_SF_VsLogT}).

The $a_\lambda$ plasma heating contributions to radiative fluxes were hypothesized as related to solar atmospheric thermodynamic profiles, i.e., DEMs ($\S$~\ref{sec:RadVsMag_BrokePLaw}). Self-consistent with this qualitative evidence for a functional $a_\lambda$ to DEM similarity, strong linear correlations were derived between the $a_\lambda$ observations to theoretical DEMs provided by the CHIANTI atomic database. In that respect, we considered that an entanglement of thermodynamic and magnetic energy contributions existed in the typical analytic approach to magnetic energy coupling assessments (i.e., Equation~\ref{eqn:PevFit}). Conclusions that we emphasize significantly contribute to the growing evidence for a marriage of dominant and diffuse background heating components, particularly their intimate connectivity. More importantly, these plasma heating components' intimacy were revealed as a relationship between magnetic energy redistribution processes and solar atmospheric thermodynamic profiles. These results, we emphasize, have far reaching applications in the fields of solar and stellar physics, as they cast new light on the utility of narrowband observations as ad hoc tools for extrapolating solar atmospheric thermodynamic profiles, and/or their accompanying photospheric LOS magnetic field strengths.

Of additional importance to this work was the evidence presented from our passband magnetic coupling analyses (i.e., Figure~\ref{fig:PLI_SF_VsLogT}) for the existence of a simple linear temperature dependence of magnetic energy redistribution from chromosphere through coronal regimes (i.e., Equation~\ref{eqn:P=TG}). However, as AIA passbands only provide moderate temperature resolution (e.g., \citealt{Guennouetal2012ApJS}), more explicit constraints of energy redistribution with temperature effects require further investigations. Nonetheless, we find confidence in Equation~\ref{eqn:P=TG} to first order approximations, as it was derived from large diverse observational data sets, i.e., they provided broad solar atmospheric spectrum coverage, (i.e., AIA passbands), and plasma conditions (i.e., CHs, QS, ARs, ARC, and FD features) across long time baselines (i.e., $\approx$\,3\,yr). Further contributing to such notions is the fact that this observationally inferred dependence of magnetic energy redistribution with temperature (Equation~\ref{eqn:P=TG}), provides the first (to our knowledge) evidence of previously speculated notions e.g., \citet{Longcope1998ApJ}; \citet{LongcopeKankelborg1999ApJ}; \citet{Alvarado-Gomez2016AA}. Thereby, the potentiality of this combined work for addressing fundamental gaps in our understanding of the role of varying thermodynamic conditions and magnetic reconnection processes, i.e., via studies of similar physical processes under differing plasma conditions, is elevated \citep{Orangeetal2014SoPh1557O}.

Our ARC results supported both high- and low-frequency energization of these features (e.g., \citealt{Warrenetal2012ApJ}), i.e., the nearly linear related radiative distributions across broad spectrum regimes with a low corona knee (Figure~\ref{fig:FvsPhi}). These observations favor ubiquitous solar atmospheric plasma heating, which, as revealed in this work, likely stem from radiative heating via multiple generation mechanisms. Therefore, we speculate that ARC heating possibly reflects runaway SOC that could be provided by coupling a diffuse background heating component such as the MHD turbulent dissipation detailed by \cite{Laming2004ApJ,Laming2009ApJ,Laming2012ApJ} with energy dissipation via SOC-like avalanches \citep{UritskyDavila2012ApJ}. It is noted; this is a hypothesized coupling which should result in significant unresolved emission from coronal fractionation as a by-product of magnetic to thermal energy conversion processes. However, as we have only statistically sampled these features from narrowband observations, it can not be ruled out that high- and low-frequency ARC plasma heating was related to the AR ages (e.g., see \citealt{SchmelzPathak2012ApJ,Dadashietal2012AA}; {\bf \citealt{Subramanianetal2014ApJ}}).

A simple theoretical coronal heating model, that fundamentally relied on this work's evidence for self-similar generation via magnetic energy redistribution across broad spectrum scales (Equation~\ref{eqn:P=TG}) was presented in $\S$~\ref{sec:CoronalHeatingApp}. In addition, discussions were presented within the framework of our model that potentially explained long withstanding issues regarding the presence of diffuse unresolved coronal emission (e.g., see \citealt{DelZannaMason2003AA}; {\bf \citealt{Subramanianetal2014ApJ}}) as a result of energetic redistribution processes. Of interest to our coronal heating theory was its alignment with that proposed by \citeauthor{Oluseyietal1999ApJ524-1105O} (\citeyear{Oluseyietal1999ApJ524-1105O}; \citeyear{Oluseyietal1999ApJ527-992O}), particularly, that it revealed self-consistent loop heating solutions with their predictions, which fell out of single dominant generation mechanism assumptions, as they speculated.

The QS energetic distribution's overlap with that of CHs (i.e., cooler atmospheric layers of Figure~\ref{fig:FvsPhi}), could point to the presence of self-similar processes leading to open-field structures, i.e., interchange reconnection events resulting in jets (e.g., \citealt{YokoyamaShibata1995Natur}). We hypothesize that similar to the finds of \cite{Orangeetal2015ApJ}, physical processes attributed to CH formation could be common in QS conditions, as the large scale magnetic field geometries lead to enhanced volumes of ``diffusely" heated plasma. Particularly that given such hypotheses favor the ubiquitous occurrence of solar atmospheric jet phenomena (e.g., \citealt{Shimojoetal1998SoPh}; \citealt{ShimojoShibata2000AdSpR}), and align with the observed QS ``clustering" of radiative to magnetic energies (Figure~\ref{fig:FvsPhi}). More specifically, these observations indicate plasma heating beyond the standard flare model \citep{Parker1963ApJS}, and suggest the inferred fundamental difference of open to closed field heating as being akin to that of jets and flares (e.g., \citealt{Shibataetal1992PASJ}; \citealt{WangSheely1993ApJ}; \citealt{Wangetal1996Sci}; \citealt{Shibataetal1997ASPC}). Thereby, our work has elevated the role of cooler atmospheric studies in elucidating the physical plasma heating processes of large scale open to closed field lines \citep{Orangeetal2015ApJ}.

Finally, this study has addressed SDO's objective to increase our understanding of the origin of solar activity \citep{Pesnelletal2012SoPh}. In addition, it has indicated proxies that hold significant potential for the field of stellar physics, mainly through providing possible means for probing distant stellar sources in currently difficult and/or undetectable energy ranges, and techniques for extrapolating radiative to magnetic field characteristics of gross feature classes via unresolved stellar disk observations, and thereby elevating SDO's extensive data archive as a tool for enhancing our understanding of stellar physics.

\section{Acknowledgements}

A portion of this work has been supported by OrangeWave Innovative Science's solar physics objective. N.B.O. acknowledges the Florida Space Grant Consortium, a NASA sponsored program administered by the University of Central Florida, grant NNX-10AM01H. N.B.O. and D.L.C. thank the University of the Virgin Islands. B.G. acknowledges financial support from NASA grant NNX13AD28A and NNX15AP95A. Any opinions, findings and conclusions or recommendations expressed in this material are those of the author(s) and do not necessarily reflect the views of NASA. The authors wish to acknowledge P.R.~Champy and M.~Patel for their assistance in the data processing and collection tasks. \\~\\

\bibliographystyle{apj}
\bibliography{MEC}

\begin{thebibliography}{117}
\expandafter\ifx\csname natexlab\endcsname\relax\def\natexlab#1{#1}\fi

\bibitem[{{Abbett}(2007)}]{Abbett2007ApJ}
{Abbett}, W.~P. 2007, \apj, 665, 1469

\bibitem[{{Achour} {et~al.}(1995){Achour}, {Brekke}, {Kjeldseth-Moe}, \&
  {Maltby}}]{Achouretal1995ApJ}
{Achour}, H., {Brekke}, P., {Kjeldseth-Moe}, O., \& {Maltby}, P. 1995, \apj,
  453, 945

\bibitem[{{Alvarado-G{\'o}mez} {et~al.}(2016){Alvarado-G{\'o}mez}, {Hussain},
  {Cohen}, {Drake}, {Garraffo}, {Grunhut}, \& {Gombosi}}]{Alvarado-Gomez2016AA}
{Alvarado-G{\'o}mez}, J.~D., {Hussain}, G.~A.~J., {Cohen}, O., {et~al.} 2016,
  \aap, 588, A28

\bibitem[{{Aschwanden} {et~al.}(2011){Aschwanden}, {Boerner}, {Schrijver}, \&
  {Malanushenko}}]{Aschwandenetal2011SolPh}
{Aschwanden}, M.~J., {Boerner}, P., {Schrijver}, C.~J., \& {Malanushenko}, A.
  2011, \solphys, 384

\bibitem[{{Aschwanden} \& {Nightingale}(2005)}]{AschwandenNightingale2005ApJ}
{Aschwanden}, M.~J., \& {Nightingale}, R.~W. 2005, \apj, 633, 499

\bibitem[{{Aschwanden} \& {Schrijver}(2002)}]{AschwandenSchrijver2002ApJS}
{Aschwanden}, M.~J., \& {Schrijver}, C.~J. 2002, \apjs, 142, 269

\bibitem[{{Bak} {et~al.}(1987){Bak}, {Tang}, \&
  {Wiesenfeld}}]{Baketal1987PhRvL}
{Bak}, P., {Tang}, C., \& {Wiesenfeld}, K. 1987, Physical Review Letters, 59,
  381

\bibitem[{{Benevolenskaya} {et~al.}(2002){Benevolenskaya}, {Kosovichev},
  {Lemen}, {Scherrer}, \& {Slater}}]{Benevolenskayaetal2002ApJ}
{Benevolenskaya}, E.~E., {Kosovichev}, A.~G., {Lemen}, J.~R., {Scherrer},
  P.~H., \& {Slater}, G.~L. 2002, \apjl, 571, L181

\bibitem[{{Berger}(1997)}]{Berger1997PhDT}
{Berger}, T.~E. 1997, PhD thesis, Standford University

\bibitem[{{Berger} {et~al.}(1997){Berger}, {Lofdahl}, {Shine}, \&
  {Title}}]{Bergeretal1997SPD}
{Berger}, T.~E., {Lofdahl}, M.~G., {Shine}, R.~A., \& {Title}, A.~M. 1997, in
  Bulletin of the American Astronomical Society, Vol.~29, AAS/Solar Physics
  Division Meeting \#28, 896

\bibitem[{{Bingert} \& {Peter}(2011)}]{BingertPeter2011AA}
{Bingert}, S., \& {Peter}, H. 2011, \aap, 530, A112

\bibitem[{{Boerner} {et~al.}(2012){Boerner}, {Edwards}, {Lemen}, {Rausch},
  {Schrijver}, {Shine}, {Shing}, {Stern}, {Tarbell}, {Title}, {Wolfson},
  {Soufli}, {Spiller}, {Gullikson}, {McKenzie}, {Windt}, {Golub}, {Podgorski},
  {Testa}, \& {Weber}}]{Boerneretal2012SoPh}
{Boerner}, P., {Edwards}, C., {Lemen}, J., {et~al.} 2012, \solphys, 275, 41

\bibitem[{{Boerner} {et~al.}(2014){Boerner}, {Testa}, {Warren}, {Weber}, \&
  {Schrijver}}]{Boerneretal2014SoPh}
{Boerner}, P.~F., {Testa}, P., {Warren}, H., {Weber}, M.~A., \& {Schrijver},
  C.~J. 2014, \solphys, 289, 2377

\bibitem[{{Bradshaw}(2003)}]{Bradshaw2003PhDT}
{Bradshaw}, S. 2003, PhD thesis, Wolfson College

\bibitem[{{Bradshaw} \& {Cargill}(2005)}]{BradshawCargill2005AA}
{Bradshaw}, S.~J., \& {Cargill}, P.~J. 2005, \aap, 437, 311

\bibitem[{{Bradshaw} \& {Cargill}(2010{\natexlab{a}})}]{BradshawCargill2010ApJ}
---. 2010{\natexlab{a}}, \apjl, 710, L39

\bibitem[{{Bradshaw} \&
  {Cargill}(2010{\natexlab{b}})}]{BradshawCargill2010ApJ717}
---. 2010{\natexlab{b}}, \apj, 717, 163

\bibitem[{{Brooks} \& {Warren}(2016)}]{BrooksWarren2016ApJ}
{Brooks}, D.~H., \& {Warren}, H.~P. 2016, \apj, 820, 63

\bibitem[{{Che} \& {Goldstein}(2014)}]{CheGoldstein2014ApJ}
{Che}, H., \& {Goldstein}, M.~L. 2014, \apjl, 795, L38

\bibitem[{{Chesny} {et~al.}(2013){Chesny}, {Oluseyi}, \&
  {Orange}}]{Chesnyetal2013ApJ}
{Chesny}, D.~L., {Oluseyi}, H.~M., \& {Orange}, N.~B. 2013, \apjl, 778, L17

\bibitem[{{Craig} \& {McClymont}(1986)}]{CraigMcClymont1986ApJ}
{Craig}, I.~J.~D., \& {McClymont}, A.~N. 1986, \apj, 307, 367

\bibitem[{{Cranmer}(2012)}]{Cranmer2012SSRv}
{Cranmer}, S.~R. 2012, \ssr, 172, 145

\bibitem[{{Dadashi} {et~al.}(2012){Dadashi}, {Teriaca}, {Tripathi}, {Solanki},
  \& {Wiegelmann}}]{Dadashietal2012AA}
{Dadashi}, N., {Teriaca}, L., {Tripathi}, D., {Solanki}, S.~K., \&
  {Wiegelmann}, T. 2012, \aap, 548, A115

\bibitem[{{Del Zanna}(2008)}]{DelZanna2008AA}
{Del Zanna}, G. 2008, \aap, 481, L49

\bibitem[{{Del Zanna}(2013)}]{DelZanna2013AA}
---. 2013, \aap, 558, A73

\bibitem[{{Del Zanna} \& {Mason}(2003)}]{DelZannaMason2003AA}
{Del Zanna}, G., \& {Mason}, H.~E. 2003, \aap, 406, 1089

\bibitem[{{Del Zanna} {et~al.}(2011){Del Zanna}, {O'Dwyer}, \&
  {Mason}}]{DelZannaetal2011AA}
{Del Zanna}, G., {O'Dwyer}, B., \& {Mason}, H.~E. 2011, \aap, 535, A46

\bibitem[{{Del Zanna} {et~al.}(2015){Del Zanna}, {Tripathi}, {Mason},
  {Subramanian}, \& {O'Dwyer}}]{DelZannaetal2015AA}
{Del Zanna}, G., {Tripathi}, D., {Mason}, H., {Subramanian}, S., \& {O'Dwyer},
  B. 2015, \aap, 573, A104

\bibitem[{{Dere}(1982)}]{Dere1982SoPh}
{Dere}, K.~P. 1982, \solphys, 77, 77

\bibitem[{{Dere} {et~al.}(1984){Dere}, {Bartoe}, \&
  {Brueckner}}]{Dereetal1984ApJ}
{Dere}, K.~P., {Bartoe}, J.-D.~F., \& {Brueckner}, G.~E. 1984, \apj, 281, 870

\bibitem[{{Dere} {et~al.}(1997){Dere}, {Landi}, {Mason}, {Monsignori Fossi}, \&
  {Young}}]{Dereetal1997AAPS}
{Dere}, K.~P., {Landi}, E., {Mason}, H.~E., {Monsignori Fossi}, B.~C., \&
  {Young}, P.~R. 1997, \aaps, 125, 149

\bibitem[{{Doschek} {et~al.}(1976){Doschek}, {Bohlin}, \&
  {Feldman}}]{Doscheketal1976ApJ}
{Doschek}, G.~A., {Bohlin}, J.~D., \& {Feldman}, U. 1976, \apjl, 205, L177

\bibitem[{{Doschek} {et~al.}(2008){Doschek}, {Warren}, {Mariska}, {Muglach},
  {Culhane}, {Hara}, \& {Watanabe}}]{Doscheketal2008ApJ}
{Doschek}, G.~A., {Warren}, H.~P., {Mariska}, J.~T., {et~al.} 2008, \apj, 686,
  1362

\bibitem[{{Fisher} {et~al.}(1985){Fisher}, {Canfield}, \&
  {McClymont}}]{Fisheretal1985ApJ}
{Fisher}, G.~H., {Canfield}, R.~C., \& {McClymont}, A.~N. 1985, \apj, 289, 414

\bibitem[{{Fisher} {et~al.}(1998){Fisher}, {Longcope}, {Metcalf}, \&
  {Pevtsov}}]{Fisheretal1998ApJ}
{Fisher}, G.~H., {Longcope}, D.~W., {Metcalf}, T.~R., \& {Pevtsov}, A.~A. 1998,
  \apj, 508, 885

\bibitem[{{Fludra} \& {Ireland}(2003)}]{FludraIreland2003AA}
{Fludra}, A., \& {Ireland}, J. 2003, \aap, 398, 297

\bibitem[{{Fontenla} {et~al.}(1990){Fontenla}, {Avrett}, \&
  {Loeser}}]{Fontenlaetal1990ApJ}
{Fontenla}, J.~M., {Avrett}, E.~H., \& {Loeser}, R. 1990, \apj, 355, 700

\bibitem[{{Fontenla} {et~al.}(1991){Fontenla}, {Avrett}, \&
  {Loeser}}]{Fontenlaetal1991ApJ}
---. 1991, \apj, 377, 712

\bibitem[{{Fontenla} {et~al.}(1993){Fontenla}, {Avrett}, \&
  {Loeser}}]{Fontenlaetal1993ApJ}
---. 1993, \apj, 406, 319

\bibitem[{{Golub} {et~al.}(1980){Golub}, {Maxson}, {Rosner}, {Vaiana}, \&
  {Serio}}]{Golubetal1980ApJ}
{Golub}, L., {Maxson}, C., {Rosner}, R., {Vaiana}, G.~S., \& {Serio}, S. 1980,
  \apj, 238, 343

\bibitem[{{Guennou} {et~al.}(2012){Guennou}, {Auch{\`e}re}, {Soubri{\'e}},
  {Bocchialini}, {Parenti}, \& {Barbey}}]{Guennouetal2012ApJS}
{Guennou}, C., {Auch{\`e}re}, F., {Soubri{\'e}}, E., {et~al.} 2012, \apjs, 203,
  26

\bibitem[{{Guerreiro} {et~al.}(2013){Guerreiro}, {Hansteen}, \& {De
  Pontieu}}]{Guerreiro2013ApJ}
{Guerreiro}, N., {Hansteen}, V., \& {De Pontieu}, B. 2013, \apj, 769, 47

\bibitem[{{Hansteen} {et~al.}(1996){Hansteen}, {Maltby}, \&
  {Malagoli}}]{Hansteenetal1996ASPC}
{Hansteen}, V., {Maltby}, P., \& {Malagoli}, A. 1996, in Astronomical Society
  of the Pacific Conference Series, Vol. 111, Astronomical Society of the
  Pacific Conference Series, ed. R.~D. {Bentley} \& J.~T. {Mariska}, 116--121

\bibitem[{{Hansteen} {et~al.}(2014){Hansteen}, {De Pontieu}, {Carlsson},
  {Lemen}, {Title}, {Boerner}, {Hurlburt}, {Tarbell}, {Wuelser}, {Pereira}, {De
  Luca}, {Golub}, {McKillop}, {Reeves}, {Saar}, {Testa}, {Tian}, {Kankelborg},
  {Jaeggli}, {Kleint}, \& {Mart{\'{\i}}nez-Sykora}}]{Hansteenetal2014Sci}
{Hansteen}, V., {De Pontieu}, B., {Carlsson}, M., {et~al.} 2014, Science, 346,
  315

\bibitem[{{Hansteen} {et~al.}(2010){Hansteen}, {Hara}, {De Pontieu}, \&
  {Carlsson}}]{Hansteenetal2010ApJ718}
{Hansteen}, V.~H., {Hara}, H., {De Pontieu}, B., \& {Carlsson}, M. 2010, \apj,
  718, 1070

\bibitem[{{Hara}(1996)}]{Hara1996PhDT}
{Hara}, H. 1996, PhD thesis, PhD thesis.~Natl.~Astronom.~Obs., Japan.164 pp.~,
  (1996)

\bibitem[{{Hollweg} \& {Yang}(1988)}]{HollwegYang1988JGR}
{Hollweg}, J.~V., \& {Yang}, G. 1988, \jgr, 93, 5423

\bibitem[{{Kano} \& {Tsuneta}(1995)}]{KanoTsuneta1995ApJ}
{Kano}, R., \& {Tsuneta}, S. 1995, \apj, 454, 934

\bibitem[{{Klimchuk}(2006)}]{Klimchuk2006SoPh}
{Klimchuk}, J.~A. 2006, \solphys, 234, 41

\bibitem[{{Klimchuk}(2014)}]{Klimchuk2014arXiv}
---. 2014, ArXiv e-prints

\bibitem[{{Klimchuk}(2015)}]{Klimchuk2015RSPTA}
---. 2015, Philosophical Transactions of the Royal Society of London Series A,
  373, 40256

\bibitem[{{Laming}(2004)}]{Laming2004ApJ}
{Laming}, J.~M. 2004, \apj, 614, 1063

\bibitem[{{Laming}(2009)}]{Laming2009ApJ}
---. 2009, \apj, 695, 954

\bibitem[{{Laming}(2012)}]{Laming2012ApJ}
---. 2012, \apj, 744, 115

\bibitem[{{Lee} \& {Magara}(2014)}]{LeeMagara2014PASJ}
{Lee}, H., \& {Magara}, T. 2014, \pasj, 66, 39

\bibitem[{{Lemen} {et~al.}(2012){Lemen}, {Title}, {Akin}, {Boerner}, {Chou},
  {Drake}, {Duncan}, {Edwards}, {Friedlaender}, {Heyman}, {Hurlburt}, {Katz},
  {Kushner}, {Levay}, {Lindgren}, {Mathur}, {McFeaters}, {Mitchell}, {Rehse},
  {Schrijver}, {Springer}, {Stern}, {Tarbell}, {Wuelser}, {Wolfson}, {Yanari},
  {Bookbinder}, {Cheimets}, {Caldwell}, {Deluca}, {Gates}, {Golub}, {Park},
  {Podgorski}, {Bush}, {Scherrer}, {Gummin}, {Smith}, {Auker}, {Jerram},
  {Pool}, {Soufli}, {Windt}, {Beardsley}, {Clapp}, {Lang}, \&
  {Waltham}}]{Lemenetal2012SolPh}
{Lemen}, J.~R., {Title}, A.~M., {Akin}, D.~J., {et~al.} 2012, \solphys, 275, 17

\bibitem[{{Li} {et~al.}(2012){Li}, {Li}, \& {Yu}}]{Lietal2012RAA}
{Li}, B., {Li}, X., \& {Yu}, H. 2012, Research in Astronomy and Astrophysics,
  12, 1693

\bibitem[{{Longcope}(1998)}]{Longcope1998ApJ}
{Longcope}, D.~W. 1998, \apj, 507, 433

\bibitem[{{Longcope} \& {Kankelborg}(1999)}]{LongcopeKankelborg1999ApJ}
{Longcope}, D.~W., \& {Kankelborg}, C.~C. 1999, \apj, 524, 483

\bibitem[{{Lu} \& {Hamilton}(1991)}]{LuHamilton1991ApJ}
{Lu}, E.~T., \& {Hamilton}, R.~J. 1991, \apjl, 380, L89

\bibitem[{{Mackay} {et~al.}(2010){Mackay}, {Karpen}, {Ballester}, {Schmieder},
  \& {Aulanier}}]{Mackayetal2010SSRv}
{Mackay}, D.~H., {Karpen}, J.~T., {Ballester}, J.~L., {Schmieder}, B., \&
  {Aulanier}, G. 2010, \ssr, 151, 333

\bibitem[{{Mandrini} {et~al.}(2000){Mandrini}, {D{\'e}moulin}, \&
  {Klimchuk}}]{Mandrinietal2000ApJ}
{Mandrini}, C.~H., {D{\'e}moulin}, P., \& {Klimchuk}, J.~A. 2000, \apj, 530,
  999

\bibitem[{{Markwardt}(2009)}]{Markwardt2009ASPC}
{Markwardt}, C.~B. 2009, in Astronomical Society of the Pacific Conference
  Series, Vol. 411, Astronomical Data Analysis Software and Systems XVIII, ed.
  D.~A. {Bohlender}, D.~{Durand}, \& P.~{Dowler}, 251

\bibitem[{{McIntosh} {et~al.}(2013){McIntosh}, {Bethge}, {Threlfall}, {De
  Moortel}, {Leamon}, \& {Tian}}]{McIntoshetal2013arXiv}
{McIntosh}, S.~W., {Bethge}, C., {Threlfall}, J., {et~al.} 2013, ArXiv e-prints

\bibitem[{{Murawski} {et~al.}(2013){Murawski}, {Ballai}, {Srivastava}, \&
  {Lee}}]{Murawskietal2013MNRAS}
{Murawski}, K., {Ballai}, I., {Srivastava}, A.~K., \& {Lee}, D. 2013, \mnras

\bibitem[{{O'Dwyer} {et~al.}(2012){O'Dwyer}, {Del Zanna}, {Badnell}, {Mason},
  \& {Storey}}]{ODwyeretal2012AA}
{O'Dwyer}, B., {Del Zanna}, G., {Badnell}, N.~R., {Mason}, H.~E., \& {Storey},
  P.~J. 2012, \aap, 537, A22

\bibitem[{{O'Dwyer} {et~al.}(2011){O'Dwyer}, {Del Zanna}, {Mason}, {Sterling},
  {Tripathi}, \& {Young}}]{ODwyeretal2011AA}
{O'Dwyer}, B., {Del Zanna}, G., {Mason}, H.~E., {et~al.} 2011, \aap, 525, A137

\bibitem[{{O'Dwyer} {et~al.}(2010){O'Dwyer}, {Del Zanna}, {Mason}, {Weber}, \&
  {Tripathi}}]{ODwyeretal2010AA}
{O'Dwyer}, B., {Del Zanna}, G., {Mason}, H.~E., {Weber}, M.~A., \& {Tripathi},
  D. 2010, \aap, 521, A21

\bibitem[{{Oluseyi} {et~al.}(1999{\natexlab{a}}){Oluseyi}, {Walker}, {Porter},
  {Hoover}, \& {Barbee}}]{Oluseyietal1999ApJ524-1105O}
{Oluseyi}, H.~M., {Walker}, II, A.~B.~C., {Porter}, J., {Hoover}, R.~B., \&
  {Barbee}, Jr., T.~W. 1999{\natexlab{a}}, \apj, 524, 1105

\bibitem[{{Oluseyi} {et~al.}(1999{\natexlab{b}}){Oluseyi}, {Walker},
  {Santiago}, {Hoover}, \& {Barbee}}]{Oluseyietal1999ApJ527-992O}
{Oluseyi}, H.~M., {Walker}, II, A.~B.~C., {Santiago}, D.~I., {Hoover}, R.~B.,
  \& {Barbee}, Jr., T.~W. 1999{\natexlab{b}}, \apj, 527, 992

\bibitem[{{Orange} {et~al.}(2013){Orange}, {Chesny}, {Hesterly}, {Patel},
  {Champey}, \& {Oluseyi}}]{Orangeetal2013ApJ}
{Orange}, N., {Chesny}, D., {Hesterly}, K., {et~al.} 2013, \apj, 778

\bibitem[{{Orange} {et~al.}(2010){Orange}, {Oluseyi}, {Bruni}, {Chesny},
  {Neira}, \& {Preuss}}]{Orangeetal2010AAS}
{Orange}, N., {Oluseyi}, H., {Bruni}, D., {et~al.} 2010, in Bulletin of the
  American Astronomical Society, Vol.~42, American Astronomical Society Meeting
  Abstracts \#215, \#422.03

\bibitem[{{Orange}(2014)}]{Orange2014PhDT}
{Orange}, N.~B. 2014, PhD thesis, Florida Institute of Technology

\bibitem[{{Orange} {et~al.}(2015){Orange}, {Chesny}, \&
  {Oluseyi}}]{Orangeetal2015ApJ}
{Orange}, N.~B., {Chesny}, D.~L., \& {Oluseyi}, H.~M. 2015, \apj, 810, 98

\bibitem[{{Orange} {et~al.}(2011){Orange}, {Oluseyi}, {Chesny}, {Neira},
  {Preuss}, {DeBoth}, {Ebert}, \& {Cohen}}]{Orangeetal2011AAS}
{Orange}, N.~B., {Oluseyi}, H.~M., {Chesny}, D., {et~al.} 2011, in Bulletin of
  the American Astronomical Society, Vol.~43, American Astronomical Society
  Meeting Abstracts \#217, \#155.10

\bibitem[{{Orange} {et~al.}(2014{\natexlab{a}}){Orange}, {Oluseyi}, {Chesny},
  {Patel}, {Champey}, {Hesterly}, {Anthony}, \&
  {Treen}}]{Orangeetal2014SoPh1901O}
{Orange}, N.~B., {Oluseyi}, H.~M., {Chesny}, D.~L., {et~al.}
  2014{\natexlab{a}}, \solphys, 289, 1901

\bibitem[{{Orange} {et~al.}(2014{\natexlab{b}}){Orange}, {Oluseyi}, {Chesny},
  {Patel}, {Hesterly}, {Preuss}, {Neira}, \&
  {Turner}}]{Orangeetal2014SoPh1557O}
---. 2014{\natexlab{b}}, \solphys, 289, 1557

\bibitem[{{Parker}(1963)}]{Parker1963ApJS}
{Parker}, E.~N. 1963, \apjs, 8, 177

\bibitem[{{Parker}(1983)}]{Parker1983GApFD}
---. 1983, Geophysical and Astrophysical Fluid Dynamics, 23, 85

\bibitem[{{Pesnell} {et~al.}(2012){Pesnell}, {Thompson}, \&
  {Chamberlin}}]{Pesnelletal2012SoPh}
{Pesnell}, W.~D., {Thompson}, B.~J., \& {Chamberlin}, P.~C. 2012, \solphys,
  275, 3

\bibitem[{{Peter} \& {Judge}(1999)}]{PeterJudge1999ApJ}
{Peter}, H., \& {Judge}, P.~G. 1999, \apj, 522, 1148

\bibitem[{{Pevtsov} {et~al.}(2003){Pevtsov}, {Fisher}, {Acton}, {Longcope},
  {Johns-Krull}, {Kankelborg}, \& {Metcalf}}]{Pevtsovetal2003ApJ}
{Pevtsov}, A.~A., {Fisher}, G.~H., {Acton}, L.~W., {et~al.} 2003, \apj, 598,
  1387

\bibitem[{{Poedts} \& {de Groof}(2004)}]{PoedtsGroof2004ESASP}
{Poedts}, S., \& {de Groof}, A. 2004, in ESA Special Publication, Vol. 575,
  SOHO 15 Coronal Heating, ed. R.~W. {Walsh}, J.~{Ireland}, D.~{Danesy}, \&
  B.~{Fleck}, 62

\bibitem[{{Press} {et~al.}(2002){Press}, {Teukolsky}, {Vetterling}, \&
  {Flannery}}]{Pressetal2002Book}
{Press}, W.~H., {Teukolsky}, S.~A., {Vetterling}, W.~T., \& {Flannery}, B.~P.
  2002, {Numerical recipes in C++ : the art of scientific computing}

\bibitem[{{Roald} {et~al.}(2000){Roald}, {Sturrock}, \&
  {Wolfson}}]{Roaldetal2000ApJ}
{Roald}, C.~B., {Sturrock}, P.~A., \& {Wolfson}, R. 2000, \apj, 538, 960

\bibitem[{{Rosner} {et~al.}(1978){Rosner}, {Tucker}, \&
  {Vaiana}}]{Rosneretal1978ApJ}
{Rosner}, R., {Tucker}, W.~H., \& {Vaiana}, G.~S. 1978, \apj, 220, 643

\bibitem[{{Saar}(1996)}]{Saar1996IAUS}
{Saar}, S.~H. 1996, in IAU Symposium, Vol. 176, Stellar Surface Structure, ed.
  K.~G. {Strassmeier} \& J.~L. {Linsky}, 237

\bibitem[{{Schmelz} {et~al.}(2013){Schmelz}, {Jenkins}, \&
  {Kimble}}]{Schmelzetal2013SoPh}
{Schmelz}, J.~T., {Jenkins}, B.~S., \& {Kimble}, J.~A. 2013, \solphys, 283, 325

\bibitem[{{Schmelz} \& {Pathak}(2012)}]{SchmelzPathak2012ApJ}
{Schmelz}, J.~T., \& {Pathak}, S. 2012, \apj, 756, 126

\bibitem[{{Schou} {et~al.}(2012){Schou}, {Scherrer}, {Bush}, {Wachter},
  {Couvidat}, {Rabello-Soares}, {Bogart}, {Hoeksema}, {Liu}, {Duvall}, {Akin},
  {Allard}, {Miles}, {Rairden}, {Shine}, {Tarbell}, {Title}, {Wolfson},
  {Elmore}, {Norton}, \& {Tomczyk}}]{Schouetal2012SoPh}
{Schou}, J., {Scherrer}, P.~H., {Bush}, R.~I., {et~al.} 2012, \solphys, 275,
  229

\bibitem[{{Schrijver}(2001)}]{Schrijver2001ApJ}
{Schrijver}, C.~J. 2001, \apj, 547, 475

\bibitem[{{Schrijver} {et~al.}(1989){Schrijver}, {Cote}, {Zwaan}, \&
  {Saar}}]{Schrijveretal1989ApJ}
{Schrijver}, C.~J., {Cote}, J., {Zwaan}, C., \& {Saar}, S.~H. 1989, \apj, 337,
  964

\bibitem[{{Shibata} {et~al.}(1997){Shibata}, {Shimojo}, {Yokoyama}, \&
  {Ohyama}}]{Shibataetal1997ASPC}
{Shibata}, K., {Shimojo}, M., {Yokoyama}, T., \& {Ohyama}, M. 1997, in
  Astronomical Society of the Pacific Conference Series, Vol. 111, Magnetic
  Reconnection in the Solar Atmosphere, ed. R.~D. {Bentley} \& J.~T. {Mariska},
  29

\bibitem[{{Shibata} {et~al.}(1992){Shibata}, {Ishido}, {Acton}, {Strong},
  {Hirayama}, {Uchida}, {McAllister}, {Matsumoto}, {Tsuneta}, {Shimizu},
  {Hara}, {Sakurai}, {Ichimoto}, {Nishino}, \& {Ogawara}}]{Shibataetal1992PASJ}
{Shibata}, K., {Ishido}, Y., {Acton}, L.~W., {et~al.} 1992, \pasj, 44, L173

\bibitem[{{Shimojo} \& {Shibata}(2000)}]{ShimojoShibata2000AdSpR}
{Shimojo}, M., \& {Shibata}, K. 2000, Advances in Space Research, 26, 449

\bibitem[{{Shimojo} {et~al.}(1998){Shimojo}, {Shibata}, \&
  {Harvey}}]{Shimojoetal1998SoPh}
{Shimojo}, M., {Shibata}, K., \& {Harvey}, K.~L. 1998, \solphys, 178, 379

\bibitem[{{Spadaro} {et~al.}(2006){Spadaro}, {Lanza}, {Karpen}, \&
  {Antiochos}}]{Spadaroetal2006ApJ}
{Spadaro}, D., {Lanza}, A.~F., {Karpen}, J.~T., \& {Antiochos}, S.~K. 2006,
  \apj, 642, 579

\bibitem[{{Subramanian} {et~al.}(2014){Subramanian}, {Tripathi}, {Klimchuk}, \&
  {Mason}}]{Subramanianetal2014ApJ}
{Subramanian}, S., {Tripathi}, D., {Klimchuk}, J.~A., \& {Mason}, H.~E. 2014,
  \apj, 795, 76

\bibitem[{{Tan}(2014)}]{Tan2014ApJ}
{Tan}, B. 2014, \apj, 795, 140

\bibitem[{{Tripathi} {et~al.}(2011){Tripathi}, {Klimchuk}, \&
  {Mason}}]{Tripathietal2011ApJ}
{Tripathi}, D., {Klimchuk}, J.~A., \& {Mason}, H.~E. 2011, \apj, 740, 111

\bibitem[{{Tripathi} {et~al.}(2012){Tripathi}, {Mason}, {Del Zanna}, \&
  {Bradshaw}}]{Tripathietal2012ApJ}
{Tripathi}, D., {Mason}, H.~E., {Del Zanna}, G., \& {Bradshaw}, S. 2012, \apjl,
  754, L4

\bibitem[{{Tripathi} {et~al.}(2009){Tripathi}, {Mason}, {Dwivedi}, {del Zanna},
  \& {Young}}]{Tripathietal2009ApJ}
{Tripathi}, D., {Mason}, H.~E., {Dwivedi}, B.~N., {del Zanna}, G., \& {Young},
  P.~R. 2009, \apj, 694, 1256

\bibitem[{{Tsuneta} {et~al.}(1991){Tsuneta}, {Acton}, {Bruner}, {Lemen},
  {Brown}, {Caravalho}, {Catura}, {Freeland}, {Jurcevich}, {Morrison},
  {Ogawara}, {Hirayama}, \& {Owens}}]{Tsunetaetal1991SoPh}
{Tsuneta}, S., {Acton}, L., {Bruner}, M., {et~al.} 1991, \solphys, 136, 37

\bibitem[{{Uritsky} \& {Davila}(2012)}]{UritskyDavila2012ApJ}
{Uritsky}, V.~M., \& {Davila}, J.~M. 2012, \apj, 748, 60

\bibitem[{{Uritsky} \& {Davila}(2014)}]{UritskyDavila2014ApJ}
---. 2014, \apj, 795, 15

\bibitem[{{Viall} \& {Klimchuk}(2012)}]{ViallKlimchuk2012ApJ}
{Viall}, N.~M., \& {Klimchuk}, J.~A. 2012, \apj, 753, 35

\bibitem[{{Vidotto} {et~al.}(2014){Vidotto}, {Gregory}, {Jardine}, {Donati},
  {Petit}, {Morin}, {Folsom}, {Bouvier}, {Cameron}, {Hussain}, {Marsden},
  {Waite}, {Fares}, {Jeffers}, \& {do Nascimento}}]{Vidottoetal2014MNRAS}
{Vidotto}, A.~A., {Gregory}, S.~G., {Jardine}, M., {et~al.} 2014, \mnras, 441,
  2361

\bibitem[{{Wang} {et~al.}(1996){Wang}, {Hawley}, \&
  {Sheeley}}]{Wangetal1996Sci}
{Wang}, Y.-M., {Hawley}, S.~H., \& {Sheeley}, Jr., N.~R. 1996, Science, 271,
  464

\bibitem[{{Wang} \& {Sheeley}(1993)}]{WangSheely1993ApJ}
{Wang}, Y.-M., \& {Sheeley}, Jr., N.~R. 1993, \apj, 414, 916

\bibitem[{{Warren} {et~al.}(2008){Warren}, {Feldman}, \&
  {Brown}}]{Warrenetal2008ApJ685}
{Warren}, H.~P., {Feldman}, U., \& {Brown}, C.~M. 2008, \apj, 685, 1277

\bibitem[{{Warren} \& {Winebarger}(2006)}]{WarrenWinebarger2006ApJ}
{Warren}, H.~P., \& {Winebarger}, A.~R. 2006, \apj, 645, 711

\bibitem[{{Warren} {et~al.}(2010){Warren}, {Winebarger}, \&
  {Brooks}}]{Warrenetal2010ApJ}
{Warren}, H.~P., {Winebarger}, A.~R., \& {Brooks}, D.~H. 2010, \apj, 711, 228

\bibitem[{{Warren} {et~al.}(2012){Warren}, {Winebarger}, \&
  {Brooks}}]{Warrenetal2012ApJ}
---. 2012, \apj, 759, 141

\bibitem[{{Winebarger} {et~al.}(2013){Winebarger}, {Tripathi}, {Mason}, \& {Del
  Zanna}}]{Winebargetal2013ApJ}
{Winebarger}, A., {Tripathi}, D., {Mason}, H.~E., \& {Del Zanna}, G. 2013,
  \apj, 767, 107

\bibitem[{{Winebarger} {et~al.}(2011){Winebarger}, {Schmelz}, {Warren}, {Saar},
  \& {Kashyap}}]{Winebargeretal2011ApJ}
{Winebarger}, A.~R., {Schmelz}, J.~T., {Warren}, H.~P., {Saar}, S.~H., \&
  {Kashyap}, V.~L. 2011, \apj, 740, 2

\bibitem[{{Wolfson} {et~al.}(2000){Wolfson}, {Roald}, {Sturrock}, \&
  {Weber}}]{Wolfsonetal2000ApJ}
{Wolfson}, R., {Roald}, C.~B., {Sturrock}, P.~A., \& {Weber}, M.~A. 2000, \apj,
  539, 995

\bibitem[{{Yokoyama} \& {Shibata}(1995)}]{YokoyamaShibata1995Natur}
{Yokoyama}, T., \& {Shibata}, K. 1995, \nat, 375, 42

\end{thebibliography}
\end{document}